\documentclass[times,twoside]{zHenriquesLab-StyleBioRxiv}

\usepackage{geometry}
\usepackage[]{multibib}
\newcites{Method}{References}

\usepackage{blindtext}
\usepackage{textcomp}

\usepackage{floatrow}
\usepackage[label font=bf,labelformat=simple]{subfig}
\usepackage{subfig}
\usepackage{float}
\usepackage{overpic}
\usepackage{bbding}
\usepackage{array}
\usepackage{makecell}
\usepackage{amsmath}
\usepackage{bookmark}
\usepackage{caption}
\DeclareCaptionLabelSeparator{pipe}{ $\boldsymbol{|}$ }
\DeclareCaptionLabelFormat{supfiglabel}{\textbf{Supplementary Fig. #2}}
\DeclareCaptionType{supplementaryfigure}[Extended Data Fig.][List of Supplementary Figures]

\usepackage{graphicx}%
\usepackage{multirow}%

\usepackage{amssymb}
\usepackage{amsfonts}%
\usepackage{mathrsfs}%
\usepackage[title]{appendix}%
\usepackage{xcolor}%
\usepackage{textcomp}%
\usepackage{manyfoot}%
\usepackage{booktabs}%
\usepackage{algorithm}%
\usepackage{algorithmicx}%
\usepackage{algpseudocode}%
\usepackage{listings}%
\usepackage{hyperref}
\usepackage{wrapfig}
\usepackage{graphics}
\usepackage{epstopdf}
\usepackage{pifont}
\leadauthor{Zhang} 
\begin{document}

\title{Generative AI  Enables Medical Image Segmentation in Ultra Low-Data Regimes}
\shorttitle{GenSeg}

\author[1]{Li~Zhang}
\author[1]{Basu~Jindal}
\author[2,3]{Ahmed~Alaa}
\author[4]{Robert~Weinreb}
\author[5]{David~Wilson}
\author[6,7]{Eran~Segal}
\author[8,9]{James~Zou}
\author[1,10  \Letter]{Pengtao~Xie}

\affil[1]{Department of Electrical and Computer Engineering, University of California San Diego, La Jolla, CA, USA}
\affil[2]{Bakar Computational Health Sciences Institute, University of California San Francisco, San Francisco, CA, USA}
\affil[3]{Department of Electrical Engineering and Computer Sciences, University of California Berkeley, Berkeley, CA, USA}
\affil[4]{Hamilton Glaucoma Center, Shiley Eye Institute, Viterbi Family Department of Ophthalmology, University of California San Diego, La Jolla, CA, USA}
\affil[5]{Division of Pulmonary, Allergy and Critical Care Medicine, Department of Medicine, University of Pittsburgh, Pittsburgh, PA, USA}
\affil[6]{Department of Computer Science and Applied Mathematics, Weizmann Institute of Science, Rehovot, Israel}
\affil[7]{Department of Molecular Cell Biology, Weizmann Institute of Science, Rehovot, Israel}
\affil[8]{Department of Biomedical Data Science, Stanford University School of Medicine, Stanford, CA, USA}
\affil[9]{Department of Computer Science, Stanford University, Stanford, CA, USA}
\affil[10]{Department of Medicine, University of California San Diego, La Jolla, CA, USA}

\maketitle

\begin{abstract}
Semantic segmentation of medical images is pivotal in applications like disease diagnosis and treatment planning. While deep learning has excelled in automating this task, a major hurdle is the need for numerous annotated segmentation masks, which are resource-intensive to produce due to the required expertise and time. This scenario often leads to ultra low-data regimes, where annotated images are extremely limited, posing significant challenges for the generalization of conventional deep learning methods on test images. To address this, we introduce a generative deep learning framework, which uniquely generates  high-quality paired segmentation masks and medical images, serving as auxiliary data for training robust models in data-scarce environments. 
Unlike traditional generative models that treat data generation and segmentation model training as separate processes, our method employs multi-level optimization for end-to-end data generation. This approach allows segmentation performance to directly influence the data generation process, ensuring that the generated data is specifically tailored to enhance the performance of the segmentation model. Our method demonstrated strong generalization performance across 9 diverse medical image segmentation tasks and on 16 datasets, in ultra-low data regimes, spanning various diseases, organs, and imaging modalities. When applied to various segmentation models, it achieved performance improvements of 10-20\% (absolute), in both same-domain and out-of-domain scenarios. Notably, it requires 8 to 20  times less training data than existing methods to achieve comparable results. This advancement  significantly improves the feasibility and cost-effectiveness of applying deep learning in medical imaging, particularly in scenarios with limited data availability. 
\end {abstract}


\begin{corrauthor}
\text{p1xie@ucsd.edu}
\end{corrauthor}

\section*{Introduction}
Medical image semantic segmentation~\cite{ronneberger2015u,isensee2021nnu,ma2024segment} is a pivotal process in the modern healthcare landscape, playing an indispensable role in diagnosing diseases~\cite{antonelli2022medical}, tracking disease progression~\cite{pu2021automated}, planning treatments~\cite{zaidi2010pet}, assisting surgeries~\cite{grammatikopoulou2021cadis}, and supporting numerous other clinical activities~\cite{peiris2023uncertainty,
wang2021annotation}.  This process involves classifying each pixel within a specific image, such as a skin dermoscopy image, with a corresponding semantic label, such as skin cancer or normal skin. 
\newline
The advent of deep learning has revolutionized this domain, offering unparalleled precision and automation in the segmentation of medical images~\cite{ronneberger2015u,chen2017deeplab,xie2021segformer,isensee2021nnu}.  Despite these advancements, training  accurate and robust deep learning models requires  extensive, annotated medical imaging datasets, which are notoriously difficult to obtain~\cite{wang2021annotation,schafer2024overcoming}. Labeling semantic segmentation masks for  medical images is both time-intensive and costly, as it necessitates annotating each pixel. It requires not only substantial human resources but also specialized domain expertise. This leads to what is termed as \textit{ultra low-data regimes} – scenarios where the availability of annotated training images is remarkably scarce. This scarcity poses a substantial challenge to the existing deep learning methodologies, causing them to overfit to training data and exhibit poor generalization performance on test images. 
\newline
\newline
To address the scarcity of labeled image-mask pairs in semantic segmentation, several strategies have been devised, including data augmentation and semi-supervised learning approaches. 
Data augmentation techniques~\cite{chen2019learning,choi2019self,sandfort2019data,nguyen2024dataset}  create synthetic pairs of images and masks, which are then utilized as supplementary training data. 
A significant limitation of these methods is that they treat data augmentation and segmentation model training as separate activities. Consequently, the process of data augmentation is not influenced by  segmentation performance, leading to a situation where the augmented data might not contribute effectively to enhancing the model's segmentation capabilities. 
Semi-supervised learning techniques~\cite{peiris2023uncertainty,ouali2020semi,mendel2020semi,chen2021semi,li2021semantic} exploit additional, unlabeled images to bolster segmentation accuracy.  Despite their potential, these methods face limitations due to the necessity for extensive volumes of unlabeled images, a requirement often difficult to fulfill in medical settings where even unlabeled data can be challenging to obtain  due to privacy issues, regulatory hurdles (e.g., IRB approvals), among others. 
\newline
\newline
Recognizing these critical gaps, we introduce a new approach - GenSeg - that leverages generative deep learning~\cite{jo2023promise,goodfellow2014generative,ho2020denoising} to  address the challenges posed by ultra low-data regimes. Our approach is capable of generating high-fidelity paired segmentation masks and medical images. This auxiliary data facilitates the training of accurate segmentation models in scenarios with extremely limited real data. What sets our approach apart from existing data generation/augmentation methods~\cite{chen2019learning,choi2019self,sandfort2019data,nguyen2024dataset} is its unique capability to facilitate end-to-end data generation through multi-level optimization~\cite{choe2023betty}.  The data generation process is intricately guided by segmentation performance, ensuring that the generated data is not only of high quality but also specifically optimized to enhance the segmentation model's performance. Furthermore, in contrast to semi-supervised segmentation tools~\cite{peiris2023uncertainty,ouali2020semi,mendel2020semi,chen2021semi,li2021semantic}, our method eliminates the need for additional unlabeled images, which are often challenging to acquire. GenSeg is a versatile, model-independent framework designed to enhance the performance of a wide range of segmentation models when integrated with them. 
\newline
\newline
GenSeg was validated  across 9  segmentation tasks on 16 datasets, covering an extensive variety of imaging modalities, diseases, and organs.
When integrated with UNet~\cite{ronneberger2015unet} and DeepLab~\cite{chen2017deeplab} in ultra low-data regimes  (for instance, with only 50 training examples), GenSeg significantly enhanced their performance, in both same-domain scenarios (where training and testing images come from the same distribution) and out-of-domain scenarios (where training and testing images originate from different distributions), achieving    performance gains of 10-20\% (absolute percentages) in most cases. GenSeg is highly data efficient, outperforming or matching the segmentation performance of baseline methods with 8-20 times fewer training examples.

\section*{Results}\label{sec2}
\begin{figure*}[t]
    \centering
    \includegraphics[width=\textwidth]{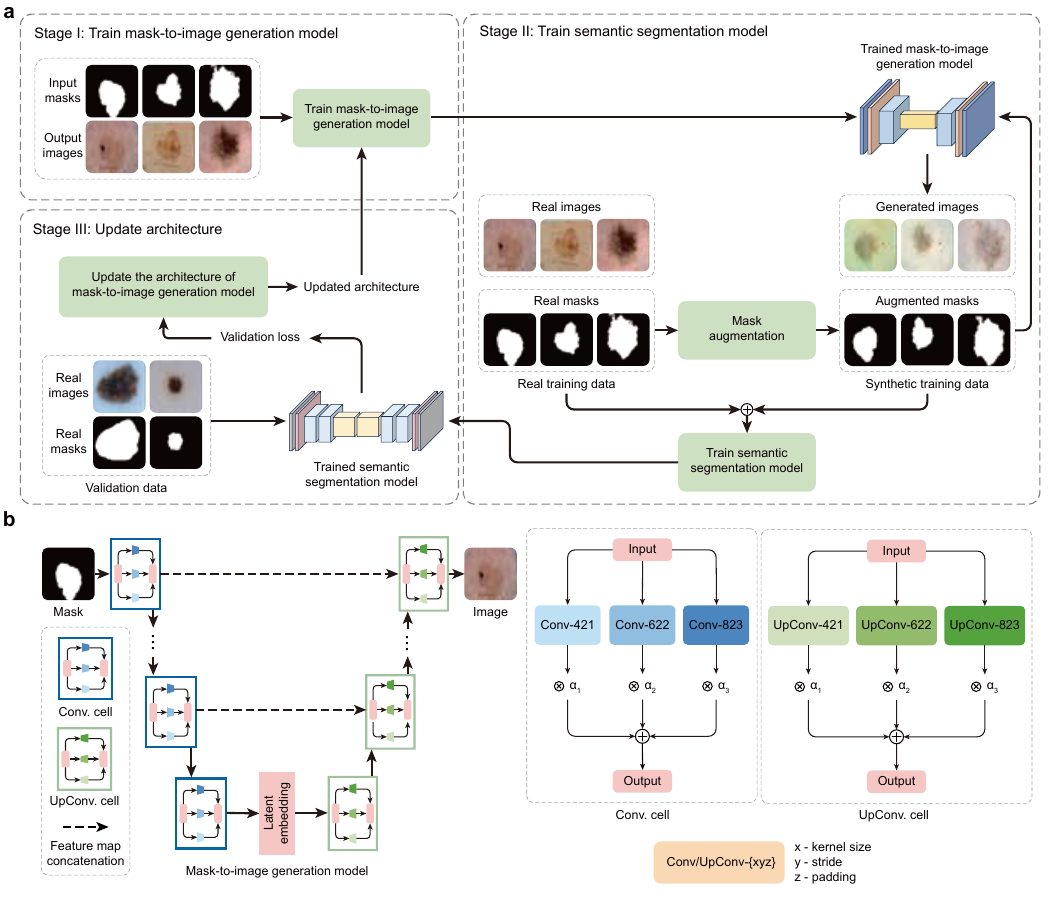}
\caption{\textbf{Proposed end-to-end data generation framework for improving medical image segmentation in ultra low-data regimes.}  \textbf{a}, Overview of the GenSeg framework.   GenSeg consists of 1) a semantic segmentation model which takes a medical image as input and predicts a segmentation mask, and 2) a mask-to-image generation model which takes a segmentation mask as input and generates a medical image. The latter  features a neural architecture that can be learned, in addition to its learnable network weights.  
GenSeg operates through three end-to-end learning stages. In {stage I}, the network weights of the mask-to-image model are trained with real mask-image pairs, while its architecture remains tentatively fixed. {Stage II} involves using the trained mask-to-image model to generate  synthetic training data. Specifically, real segmentation masks undergo augmentation procedures to produce augmented masks which are then inputted into the mask-to-image model to generate corresponding images. These images, paired with the augmented masks, are used to train the semantic segmentation model, alongside real data. In {stage III}, the trained segmentation model is evaluated on a real validation dataset, and the resulting validation loss - which reflects the performance of the mask-to-image model's architecture - is used to update this architecture. Following this update, the model re-enters Stage I for further training, and this cycle continues until convergence. \textbf{b}, Searchable architecture of the mask-to-image generation model. It comprises an encoder and a decoder. The encoder processes an input mask into a latent representation using a series of searchable convolution (Conv.) cells. The decoder employs a stack of searchable up-convolution (UpConv.) cells to convert the latent representation back into an output medical image. Each cell contains multiple candidate operations characterized by varying kernel sizes, strides, and padding options. Each operation is associated with a weight $\alpha$ denoting its importance. The process of architecture search involves optimizing these importance weights. After the learning phase, only the candidate operations with the highest weights are incorporated into the final model architecture.}

    \label{fig:overview}
    
\end{figure*}

\subsection*{GenSeg overview} GenSeg is an end-to-end data generation framework designed to generate  high-quality, labeled data, to enable the training of accurate medical image segmentation models in ultra low-data regimes (Fig.~\ref{fig:overview}a). 
Our framework integrates two components: a data generation model and a semantic segmentation model. The data generation model is responsible for generating synthetic pairs of medical images and their corresponding segmentation masks. This generated data serves as the training material for the segmentation model. In our data generation process, we introduce a  reverse generation mechanism. This mechanism initially generates segmentation masks, and subsequently, medical images, adhering to a progression from simpler to more complex tasks. Specifically, given an  expert-annotated real segmentation mask, we apply basic image augmentation operations to produce an augmented mask, which is then inputted into a deep generative model to generate the corresponding medical image. A key distinction of our method lies in the architecture of this generative model. Unlike traditional models~\cite{goodfellow2014generative,brock2018large,ho2020denoising,song2020score} that rely on manually designed architecture, our model automatically learns this architecture from data (Fig.~\ref{fig:overview}b). This adaptive architecture enables more nuanced and effective generation of medical images, tailored to the specific characteristics of the augmented segmentation masks.
\newline
\newline
GenSeg features an end-to-end data generation strategy, which ensures a synergistic relationship between the generation of data and the performance of the segmentation model. By closely aligning the data generation process with the needs and feedback of the segmentation model, GenSeg ensures the relevance and utility of the generated data for effective  training of the segmentation model. To evaluate the effectiveness of the generated data, we first train a semantic segmentation model using this data. We then assess the model's performance on a validation set consisting of real medical images, each accompanied by an expert-annotated segmentation mask.  
The model's validation performance serves as a reflection of the quality of the generated data: if the data is of low quality, the segmentation model trained on it will show poor performance during validation. By concentrating on improving the model's validation performance, we can, in turn, enhance the quality of the generated data. 
\newline
\newline
Our approach utilizes a multi-level optimization (MLO)~\cite{choe2023betty} strategy to achieve end-to-end data generation. MLO involves a series of nested optimization problems, where the optimal parameters from one level serve as inputs for the objective function at the next level. Conversely, parameters that are not yet optimized at a higher level are fed back as inputs to lower levels. This yields a dynamic, iterative process that solves optimization problems in different levels jointly. 
Our method employs a three-tiered MLO process, executed end-to-end. The first level focuses on training the weight parameters of our data generation model, while keeping its learnable architecture constant. At the second level, this trained model is used to produce synthetic image-mask pairs, which are then employed to train a semantic segmentation model. The final level involves validating the segmentation model using real medical images with expert-annotated masks. The performance of the segmentation model in this validation phase is a function of the architecture of the generation model. We optimize this architecture by minimizing the validation loss.  
By jointly solving the three levels of nested optimization problems, we can concurrently train data generation and semantic segmentation models in an end-to-end manner. 
\newline
\newline
Our framework was validated  for a variety of medical imaging segmentation tasks across 16 datasets, 
spanning a diverse spectrum of imaging techniques, diseases, lesions, and organs. 
These tasks comprise segmentation of skin lesions from dermoscopy images, breast cancer from ultrasound images, placental vessels from fetoscopic images, polyps from colonoscopy images, foot ulcers from standard camera images, intraretinal cystoid fluid from optical coherence tomography (OCT) images,  lungs from chest X-ray images, and left ventricles and myocardial wall  from echocardiography images.

\begin{figure*}
    \centering
    \includegraphics[]{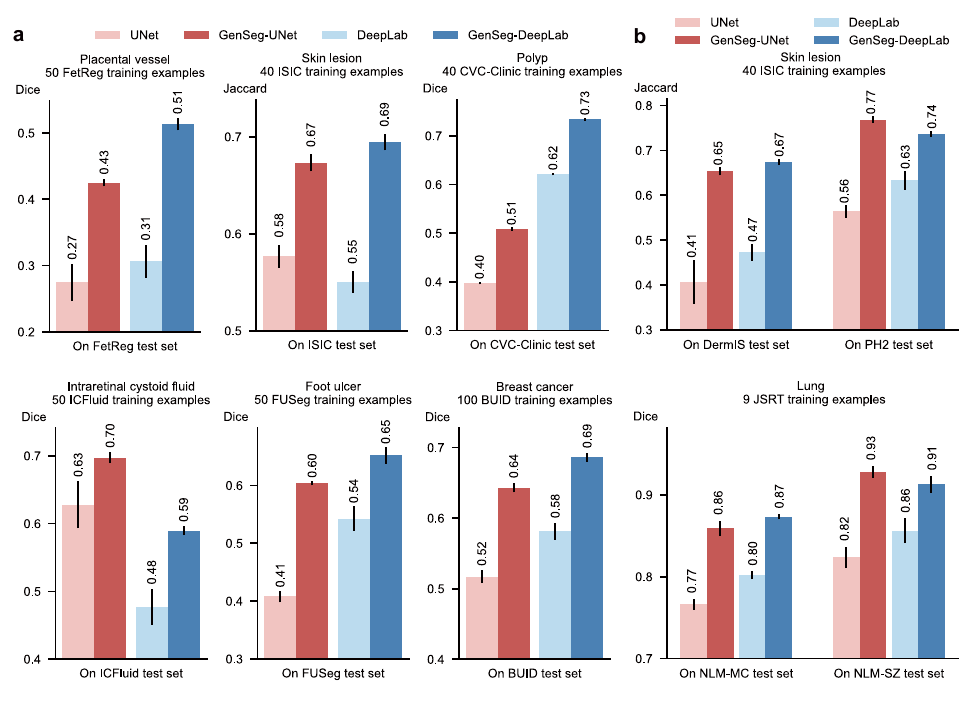}
    \caption{\textbf{GenSeg significantly boosted both in-domain and out-of-domain generalization performance, particularly in ultra low-data regimes}.  \textbf{a},  The performance of GenSeg applied to UNet (GenSeg-UNet) and DeepLab (GenSeg-DeepLab) under in-domain settings (test and training data are from the same domain) in the tasks of segmenting placental vessels, skin lesions, polyps, intraretinal cystoid fluids, foot ulcers, and breast cancer using extremely limited training data (50, 40, 40, 50, 50, and 100 examples from the FetReg, ISIC, CVC-Clinic, ICFluid, FUSeg, and BUID datasets,  respectively for each task), compared to vanilla UNet and DeepLab. \textbf{b}, The performance of GenSeg-UNet and GenSeg-DeepLab under out-of-domain settings (test and training data are from different domains) in segmenting skin lesions (using only 40 examples from the ISIC dataset for training, and the DermIS and PH2 datasets for testing) and lungs (using only 9 examples from the JSRT dataset for training, and the NLM-MC and NLM-SZ datasets for testing), compared to vanilla UNet and DeepLab.} 
    \label{fig:main-results}
\end{figure*}

\begin{figure*}
    \centering
    \includegraphics[]{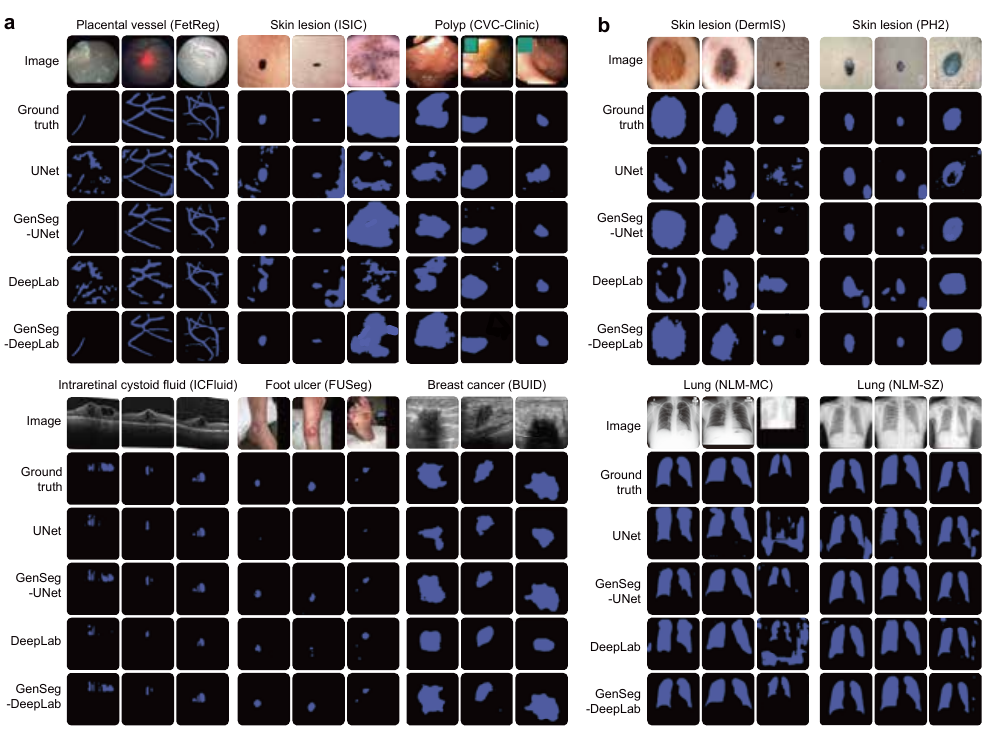}
    \caption{\textbf{GenSeg improves in-domain and out-of-domain generalization performance across a variety of segmentation tasks covering diverse diseases, organs, and imaging modalities.} \textbf{a}, Visualizations of segmentation masks predicted by GenSeg-DeepLab and GenSeg-UNet under in-domain settings in the tasks of segmenting placental vessels, skin lesions, polyps, intraretinal cystoid fluids, foot ulcers, and breast cancer using extremely limited training data (50, 40, 40, 50, 50, and 100 examples from the FetReg, ISIC, CVC-Clinic, ICFluid, FUSeg, and BUID datasets), compared to vanilla UNet and DeepLab.  \textbf{b}, Visualizations of segmentation masks predicted by GenSeg-DeepLab and GenSeg-UNet under out-of-domain settings in segmenting skin lesions (using only 40 examples from the ISIC dataset for training, and the DermIS and PH2 datasets for testing) and lungs (using only 9 examples from the JSRT dataset for training, and the NLM-MC and NLM-SZ datasets for testing), compared to vanilla UNet and DeepLab.}
    \label{fig:quali}
\end{figure*}

\subsection*{GenSeg enables accurate  segmentation  in ultra low-data regimes}

We evaluated GenSeg's performance in  ultra low-data regimes. Our method involved three-fold cross-validation on each dataset. GenSeg, being a versatile framework, facilitates training various backbone segmentation models with its generated data. To demonstrate this versatility, we applied GenSeg to two popular models: UNet~\cite{ronneberger2015unet} and DeepLab~\cite{chen2017deeplab}, resulting in GenSeg-UNet and GenSeg-DeepLab, respectively. 
GenSeg-DeepLab and GenSeg-UNet demonstrated  significant performance improvements over DeepLab and UNet in scenarios with extremely limited data (Fig.~\ref{fig:main-results}a and Extended Data Fig.~\ref{fig:swin}b). Specifically, in the tasks of segmenting placental vessels, skin lesions, polyps, intraretinal cystoid fluids, foot ulcers, and breast cancer, with training sets as small as 50, 40, 40, 50, 50, and 100 samples respectively, GenSeg-DeepLab outperformed DeepLab substantially, with absolute percentage gains of 20.6\%, 14.5\%, 11.3\%, 11.3\%, 10.9\%, and 10.4\%. 
Similarly, GenSeg-UNet surpassed UNet by significant margins, recording absolute percentage improvements of 15\%, 9.6\%, 11\%, 6.9\%, 19\%, and 12.6\% across these tasks. The extremely limited size of these training datasets presents significant challenges for accurately training DeepLab and UNet models.  For example, DeepLab's effectiveness in these tasks is limited, with performance varying from 0.31 to 0.62, averaging 0.51. In contrast, using our method, the performance significantly improves, ranging from 0.51 to 0.73 and averaging 0.64. This highlights the strong capability of our approach to achieve precise segmentation in ultra low-data regimes. 
Moreover, these segmentation tasks are highly diverse. For example,  placental vessels involve complex branching structures, skin lesions vary in shape and size, and polyps require differentiation from surrounding mucosal tissue.  GenSeg demonstrated robust performance enhancements across these diverse tasks, underscoring its strong capability in achieving accurate segmentation across different diseases, organs, and imaging modalities.

\subsection*{GenSeg enables robust generalization in  out-of-domain settings}

\begin{figure*}
    \centering
    \includegraphics{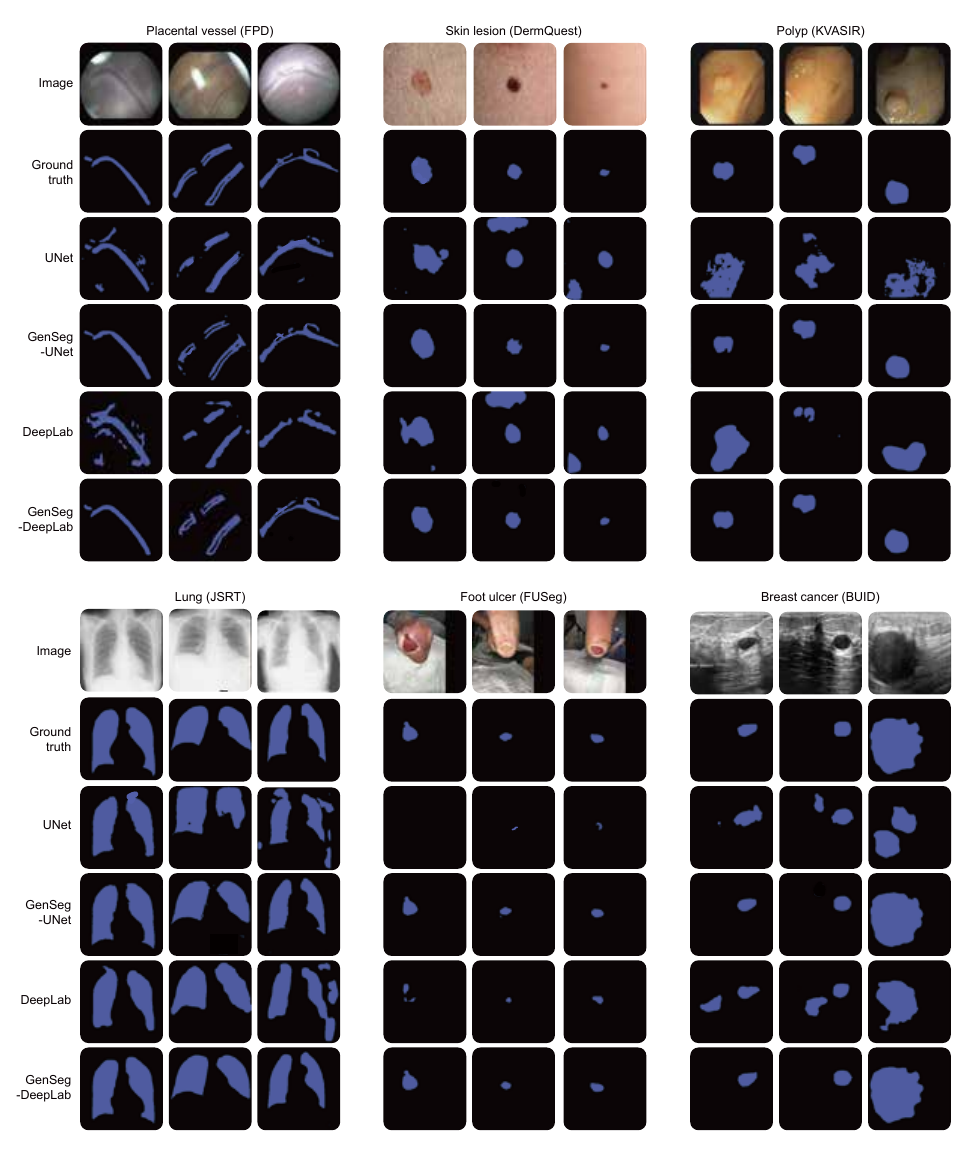}
    \captionof{supplementaryfigure}{\textbf{Additional visualizations of predicted segmentation masks.} 
    }
    \label{fig:qua_appen1}
\end{figure*}

\begin{figure*}
    \centering
    \includegraphics{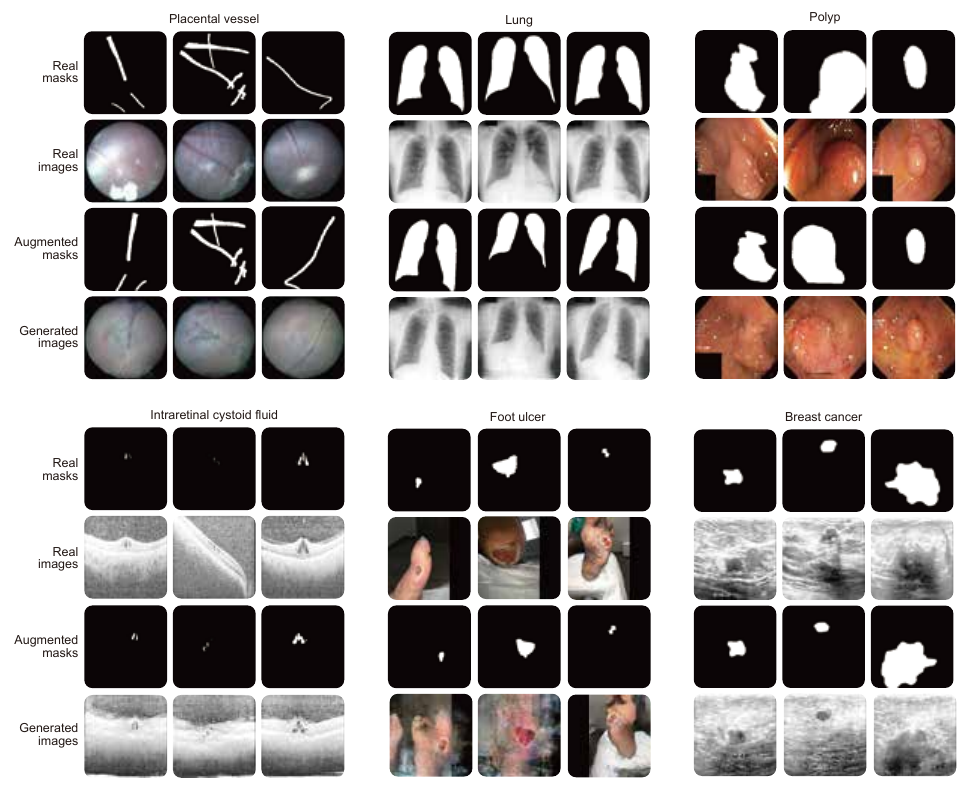}
    \captionof{supplementaryfigure}{\textbf{Visualizations of image-mask pairs generated by GenSeg.} 
    Synthetic segmentation masks and medical images  generated by GenSeg in tasks of segmenting placental vessels, lungs, polyps, intraretinal cystoid fluid, foot ulcers, and breast cancer. 
    }
    \label{fig:qua_appen2}
\end{figure*}

Besides in-domain evaluation where the test and training images were from disjoint subsets of the same dataset, we also evaluated GenSeg's effectiveness in out-of-domain (OOD) scenarios, wherein the training and test images originate from distinct datasets. The OOD evaluations were also conducted in ultra low-data regimes, where the number of training examples was restricted to only 9 or 40. Our evaluations focused on two segmentation tasks: the segmentation of skin lesions from dermoscopy images and the segmentation of lungs from chest X-rays. For the task of skin lesion segmentation, we trained our models using 40 examples from the ISIC dataset. These models were then tested on two external datasets,  DermIS and PH2, to evaluate their performance outside the ISIC domain. In the lung segmentation task, we utilized 9 training examples from the JSRT dataset and conducted evaluations on two additional datasets, NLM-SZ and NLM-MC, to test the models' adaptability beyond the JSRT domain. GenSeg showed superior out-of-domain generalization capabilities (Fig.~\ref{fig:main-results}b). In skin lesion segmentation, GenSeg-UNet substantially outperformed UNet, achieving a Jaccard index of 0.65 compared to UNet's 0.41 on the DermIS dataset, and 0.77 versus 0.56 on  PH2. Similarly, in lung segmentation, GenSeg-UNet demonstrated superior performance with a Dice score of 0.86 compared to UNet's 0.77 on NLM-MC, and  0.93 against 0.82 on NLM-SZ. Similarly, GenSeg-DeepLab significantly outperformed  DeepLab:  it achieved 0.67 compared to 0.47 on DermIS, 0.74 vs. 0.63 on PH2, 0.87 vs. 0.80 on NLM-MC, and 
0.91 vs. 0.86 on NLM-SZ.  Fig.~\ref{fig:quali} and Extended Data Fig. \ref{fig:qua_appen1} visualize some randomly selected segmentation examples. Both GenSeg-UNet and GenSeg-DeepLab accurately segmented a wide range of disease targets and organs across various imaging modalities with their predicted masks closely resembling the ground truth, under both in-domain (Fig.~\ref{fig:quali}a and Extended Data Fig. \ref{fig:qua_appen1}) and out-of-domain (Fig. \ref{fig:quali}b) settings.  In contrast, UNet and DeepLab struggled to achieve similar levels of accuracy, often producing masks that were less precise and exhibited inconsistencies in complex anatomical regions. This disparity underscores the advanced capabilities of  GenSeg  in handling varied and challenging segmentation tasks. 
Extended Data Fig. \ref{fig:qua_appen2}  presents several mask-image pairs generated by GenSeg. The generated images not only exhibit a high degree of realism but also demonstrate excellent semantic alignment with their corresponding masks.

\begin{figure*}
    \centering
    \includegraphics[]{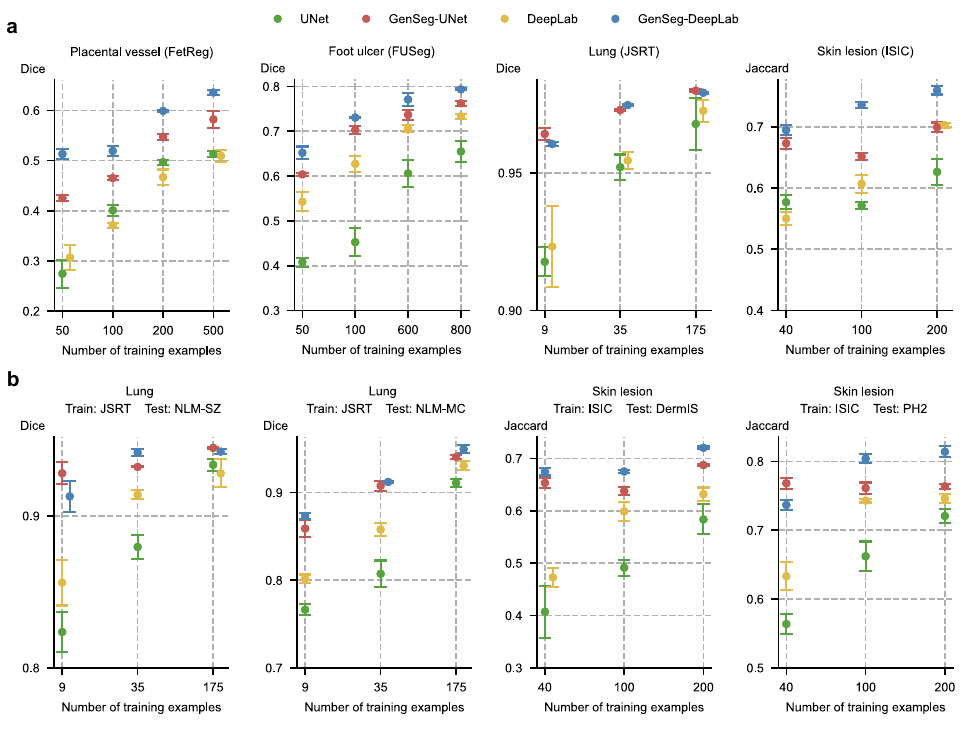}
    \caption{\textbf{GenSeg achieves performance on par with baseline models while requiring significantly fewer training examples.} \textbf{a}, The in-domain generalization performance of GenSeg-UNet and GenSeg-DeepLab with different numbers of training examples from the FetReg, FUSeg, JSRT, and ISIC datasets 
    in segmenting placental vessels, foot ulcers, lungs, and skin lesions, compared to UNet and DeepLab. \textbf{b}, The out-of-domain generalization performance of GenSeg-UNet and GenSeg-DeepLab with different numbers  of training examples in segmenting lungs (using examples from JSRT for training, and NLM-SZ and NLM-MC for testing) and skin lesions (using examples from ISIC for training, and DermIS and PH2 for testing), compared to UNet and DeepLab.}
    \label{fig:result2}
\end{figure*}

\subsection*{GenSeg achieves comparable performance to baselines with significantly fewer training examples}

In comparing the number of training examples required for GenSeg and baseline models to achieve similar performance, GenSeg consistently required fewer examples. Fig.~\ref{fig:result2} illustrates this point by plotting segmentation performance (y-axis) against the number of training examples (x-axis) for various methods. Methods that are closer to the upper left corner of the subfigure are considered more sample-efficient, as they achieve superior segmentation performance with fewer training examples. 
Across all subfigures, our methods consistently position nearer to these optimal upper left corners compared to the baseline methods. First,  GenSeg  demonstrates superior  sample-efficiency under  in-domain  settings (Fig.~\ref{fig:result2}a).  
For example, 
in the placental vessel segmentation task, GenSeg-DeepLab achieved a  Dice score of 0.51 with only 50 training examples, a ten-fold reduction compared to DeepLab's 500 examples needed to reach the same score.  
In foot ulcer segmentation, to reach a  Dice score around 0.6, UNet needed 600 examples, in contrast to GenSeg-UNet which required only 50 examples, a twelve-fold reduction. DeepLab required 800 training examples for a Dice score of 0.73, whereas GenSeg-DeepLab achieved the same score with only 100 examples, an eight-fold reduction. 
In lung segmentation, achieving a Dice score of  0.97 required 175 examples for UNet, whereas GenSeg-UNet needed just 9 examples, representing a 19-fold reduction.
Second, the sample  efficiency of GenSeg is also evident in out-of-domain (OOD) settings (Fig.~\ref{fig:result2}b).
For example, in lung segmentation, achieving an OOD generalization performance of 0.93 on the NLM-SZ dataset required 175 training examples from the JSRT dataset for UNet, while GenSeg-UNet needed only 9 examples, representing a 19-fold reduction. In skin lesion segmentation, GenSeg-DeepLab, trained with only 40 ISIC examples, reached a Jaccard index of 0.67 on DermIS, a performance that DeepLab could not match even with 200 examples.

\begin{figure*}
    \centering
    \includegraphics[]{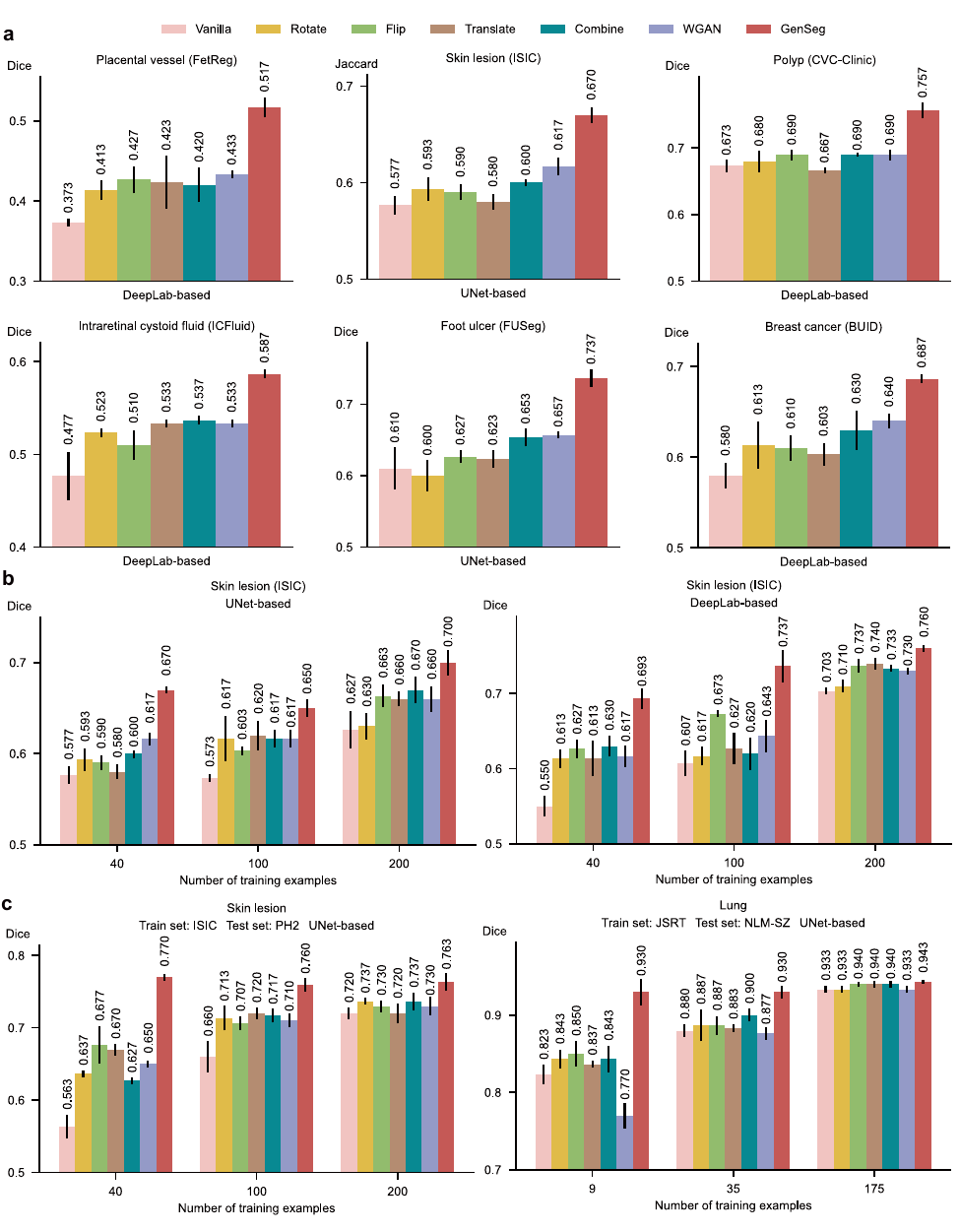}
    \caption{\textbf{GenSeg significantly outperformed widely used data augmentation and generation methods}. 
    \textbf{a},  GenSeg's in-domain generalization performance compared to baseline methods including Rotate, Flip, Translate, Combine, and WGAN, when used with UNet or DeepLab in segmenting 
    placental vessels, skin lesions, polyps, intraretinal cystoid fluids, foot ulcers, and breast cancer 
    using the FetReg, ISIC, CVC-Clinic, ICFluid, FUSeg, and BUID datasets.  
    \textbf{b}, GenSeg's in-domain generalization performance compared to baseline methods using a varying number of training examples from the ISIC dataset for segmenting skin lesions, with UNet and DeepLab as the backbone segmentation models. \textbf{c}, GenSeg's out-of-domain generalization performance compared to baseline methods across varying numbers of training examples in segmenting lungs (using examples from JSRT  for training, and  NLM-SZ and NLM-MC  for testing) and skin lesions (using examples from  ISIC  for training, and  DermIS and PH2  for testing), with UNet and DeepLab as the backbone segmentation models.
    }
    \label{fig:aug}
\end{figure*}

\begin{figure*}
    \centering
    \includegraphics[]{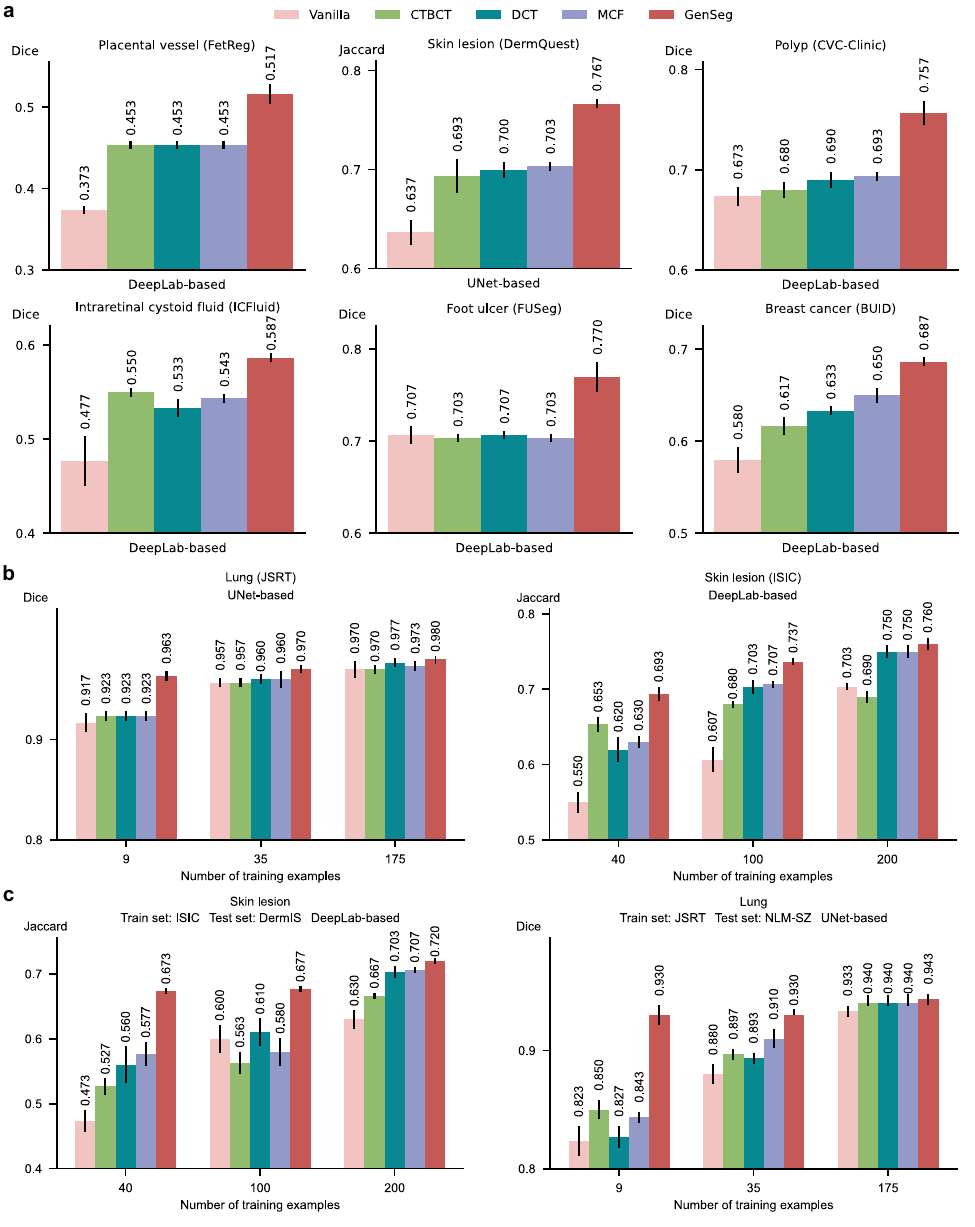}
    \caption{\textbf{GenSeg significantly outperformed state-of-the-art semi-supervised segmentation methods.} 
    \textbf{a},  GenSeg's in-domain generalization performance compared to baseline methods including CTBCT, DCT, and MCF, when used with UNet or DeepLab in segmenting placental vessels, skin lesions, polyps, intraretinal cystoid fluids, foot ulcers, and breast cancer  utilizing 
    the FetReg, DermQuest, CVC-Clinic, ICFluid, FUSeg, and BUID datasets.  
    \textbf{b}, GenSeg's in-domain generalization performance compared to baseline methods using a varying number of training examples from the ISIC and JSRT datasets for segmenting skin lesions and lungs, with UNet and DeepLab as the backbone segmentation models. \textbf{c}, GenSeg's out-of-domain generalization performance compared to baseline methods across varying numbers of training examples in segmenting lungs (using examples from JSRT  for training, and  NLM-SZ and NLM-MC  for testing) and skin lesions (using examples from  ISIC  for training, and  DermIS and PH2  for testing), with UNet and DeepLab as the backbone segmentation models.}
    \label{fig:semi}
\end{figure*}

\begin{figure*}
    \centering
    \includegraphics[]{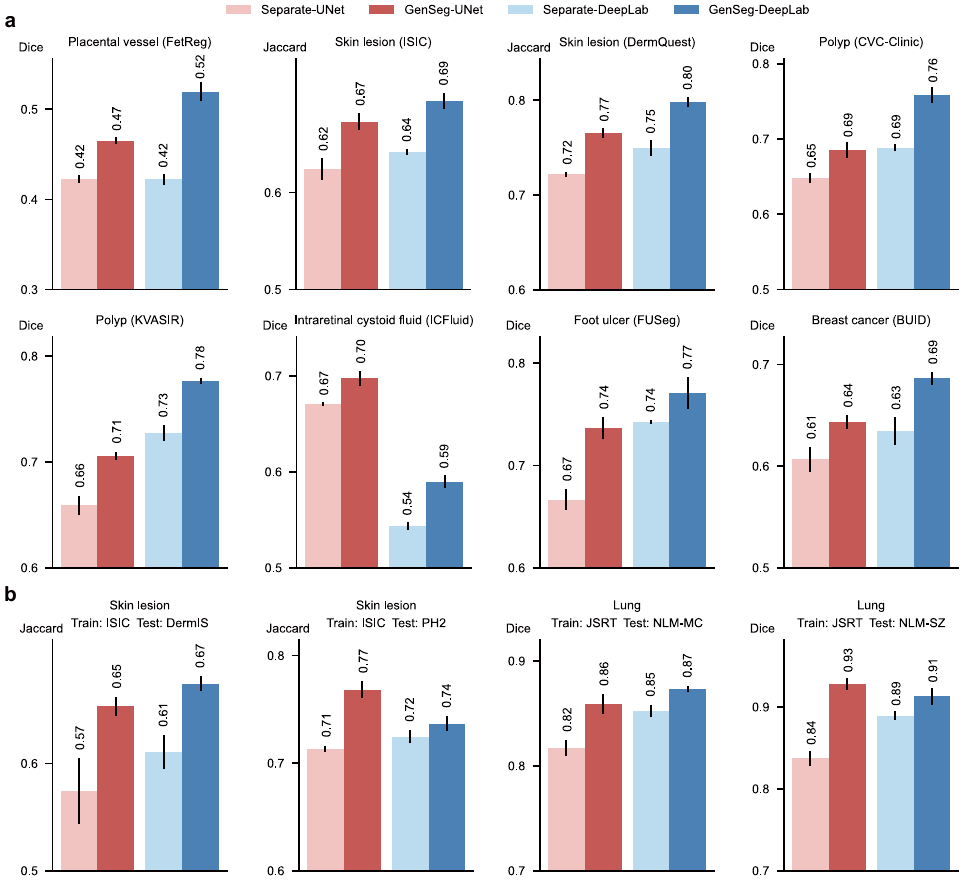}
    \caption{\textbf{GenSeg's end-to-end data generation mechanism significantly outperformed baselines'  separate generation mechanism.} \textbf{a}, The in-domain generalization performance of GenSeg which performs data generation and segmentation model training end-to-end, compared to the Separate baseline which performs the two processes separately, when used with UNet or DeepLab in segmenting placental vessels, skin lesions, polyps, intraretinal cystoid fluids, foot ulcers, and breast cancer  utilizing the FetReg, ISIC, DermQuest, CVC-Clinic, KVASIR, ICFluid, FUSeg, and BUID datasets. \textbf{b}, GenSeg's out-of-domain generalization performance compared to the Separate baseline in segmenting skin lesions (using examples from ISIC  for training, and  DermIS and PH2  for testing) and lungs (using examples from JSRT  for training, and  NLM-SZ and NLM-MC  for testing), with UNet and DeepLab as the backbone segmentation models.}
    \label{fig:resultend-to-end}
\end{figure*}

\subsection*{GenSeg outperforms widely used data augmentation and generation tools}

\begin{figure*}
    \centering
    \includegraphics{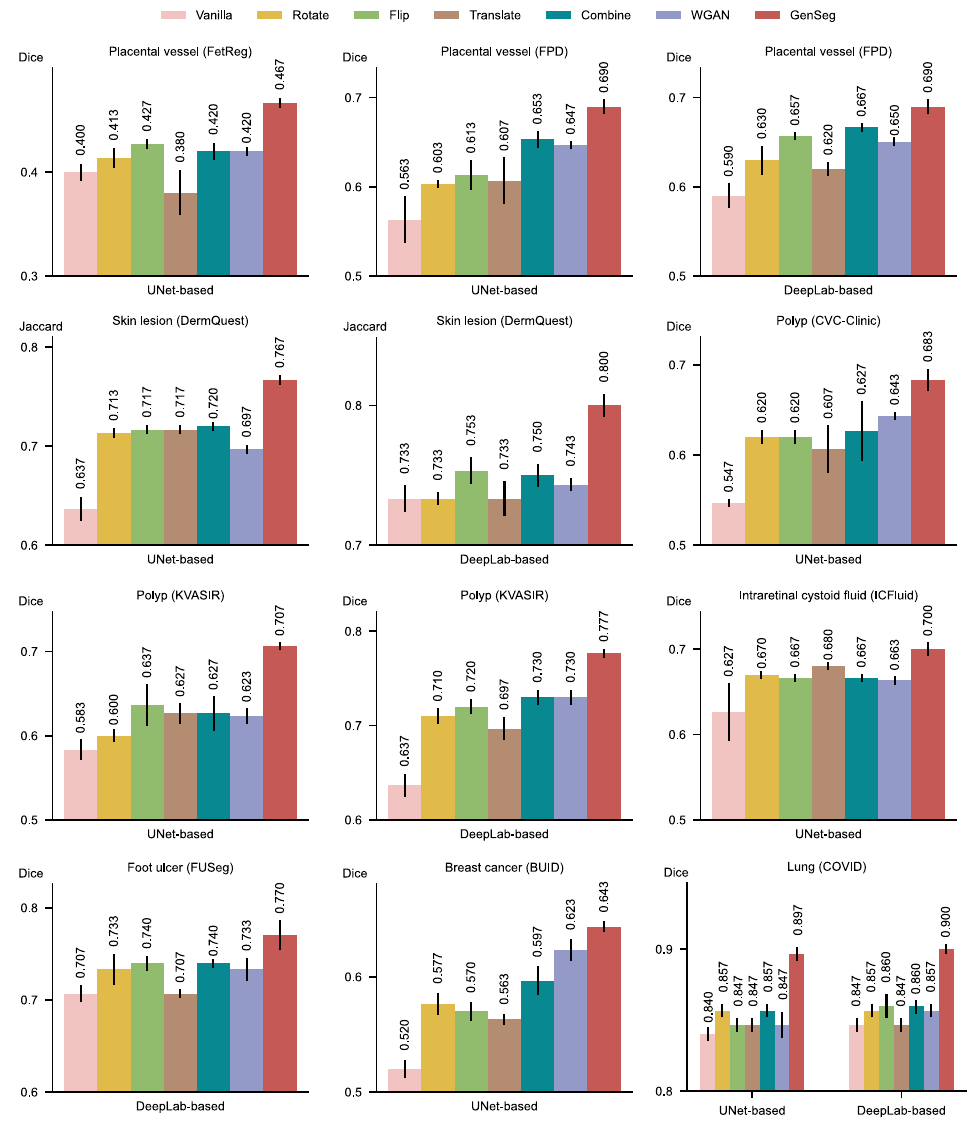}
    \captionof{supplementaryfigure}{\textbf{Further comparison of GegSeg with data augmentation and generation methods.}
    GenSeg’s in-domain generalization performance compared
to baseline methods including Rotate, Flip, Translate, Combine, and WGAN, when used with UNet or DeepLab in segmenting placental vessels, skin lesions, polyps,
intraretinal cystoid fluids, foot ulcers,  breast cancer, and lungs,  using the FetReg, FPD, DermQuest, CVC-Clinic, KVASIR, ICFluid, FUSeg, BUID, and COVID datasets.}
    \label{fig:aug_single}
\end{figure*}

\begin{figure*}
    \centering
    \includegraphics{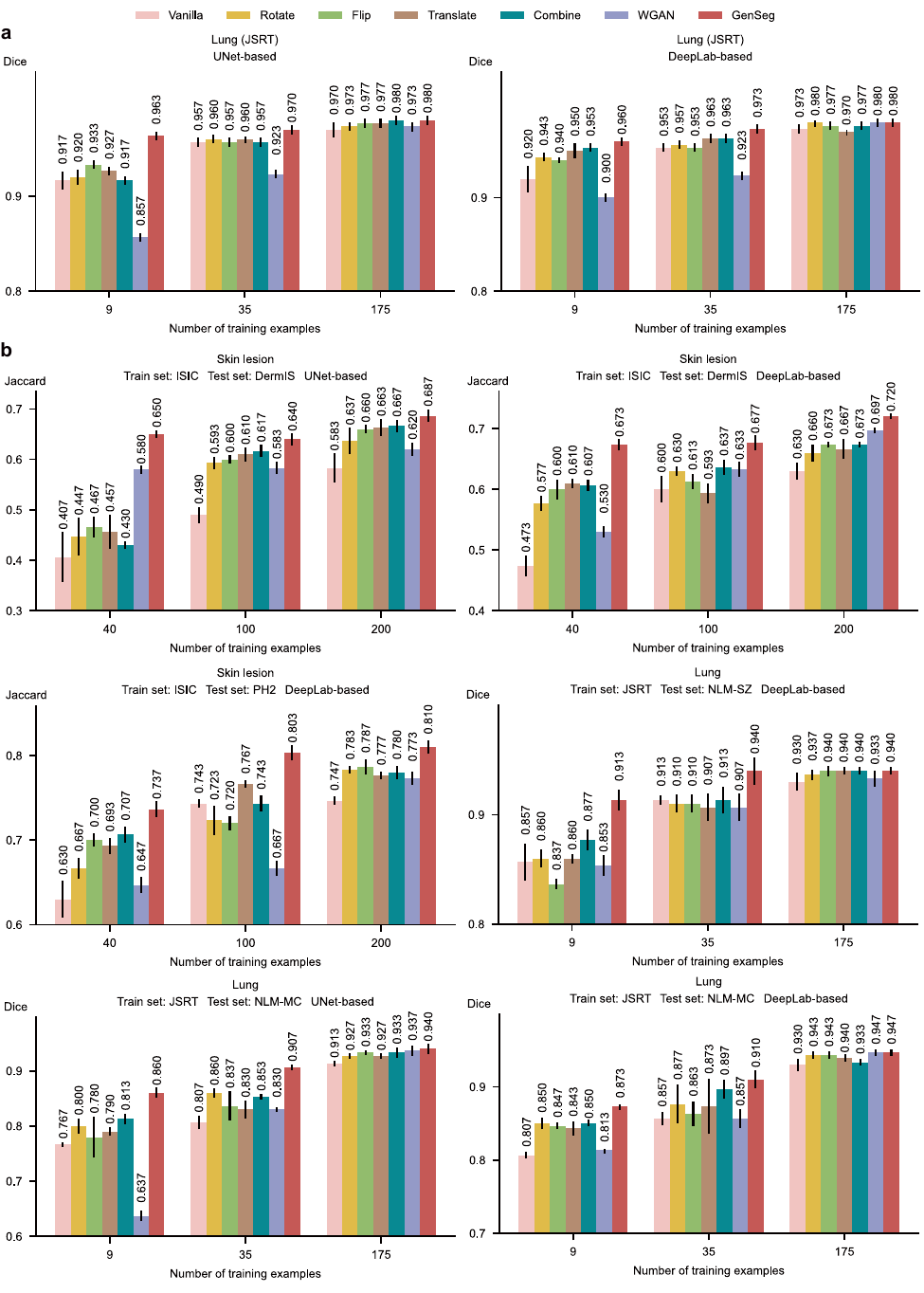}
    \vspace{-0.4cm}
    \captionof{supplementaryfigure}{\textbf{
    Further comparison of GegSeg with data augmentation and generation methods across varying numbers of training examples.     
    } \textbf{a,} Comparison of in-domain generalization performance 
    for lung segmentation using the JSRT dataset. 
     \textbf{b}, Comparison of out-of-domain generalization performance in segmenting skin lesions (using the ISIC dataset for training, DermIS and PH2 for testing) and lungs (using JSRT for training, NLM-SZ and NLM-MC for testing).}
    \label{fig:aug_varying}
\end{figure*}

We compared GenSeg against prevalent data augmentation methods, including rotation, flipping, and translation, as well as their combinations.   Furthermore, GenSeg was benchmarked against a data generation approach~\cite{neff2018generative}, which is based on the Wasserstein Generative Adversarial Network (WGAN)~\cite{arjovsky2017wasserstein}. 
For each baseline augmentation method, the same hyperparameters (e.g., rotation angle) were consistently applied to both the input image and the corresponding output mask within each training example, resulting in augmented image-mask pairs.  
GenSeg significantly  surpassed these methods  under in-domain settings (Fig.~\ref{fig:aug}a  and Extended Data Fig.~\ref{fig:aug_single}). 
For instance, in foot ulcer segmentation  using  UNet as the backbone segmentation model, GenSeg attained a Dice  score of 0.74, significantly  surpassing the top baseline method, WGAN, which achieved 0.66. Similarly, in polyp  segmentation with DeepLab, GenSeg scored 0.76, significantly outperforming the best baselines - Flip, Combine, and WGAN - which scored 0.69. 
GenSeg also demonstrated superior out-of-domain (OOD) generalization performance compared to the baselines (Fig.~\ref{fig:aug}c and Extended Data Fig. \ref{fig:aug_varying}b). For instance, in UNet-based skin lesion segmentation,  with  40 training examples from the ISIC dataset, GenSeg achieved a Dice score of 0.77 on the PH2 dataset, substantially surpassing  the best-performing baseline, Flip, which scored 0.68. Moreover, GenSeg demonstrated comparable performance to baseline methods  with fewer training examples (Fig.~\ref{fig:aug}b and Extended Data Fig. \ref{fig:aug_varying}a) under in-domain settings. For instance, using only 40 training examples for skin lesion segmentation with UNet, GenSeg achieved a Dice score of 0.67. In contrast, the best performing baseline, Combine, required 200 examples to reach the same score. Similarly,  with  fewer training examples, GenSeg achieved comparable performance to baseline methods under out-of-domain settings (Fig.~\ref{fig:aug}c and Extended Data Fig. \ref{fig:aug_varying}b).  
For example, in lung  segmentation with UNet, GenSeg reached a Dice score of 0.93 using just 9 training examples, whereas the best performing baseline required 175 examples to achieve a similar score.

\subsection*{GenSeg outperforms state-of-the-art semi-supervised segmentation  methods}

\begin{figure*}
    \centering
    \includegraphics{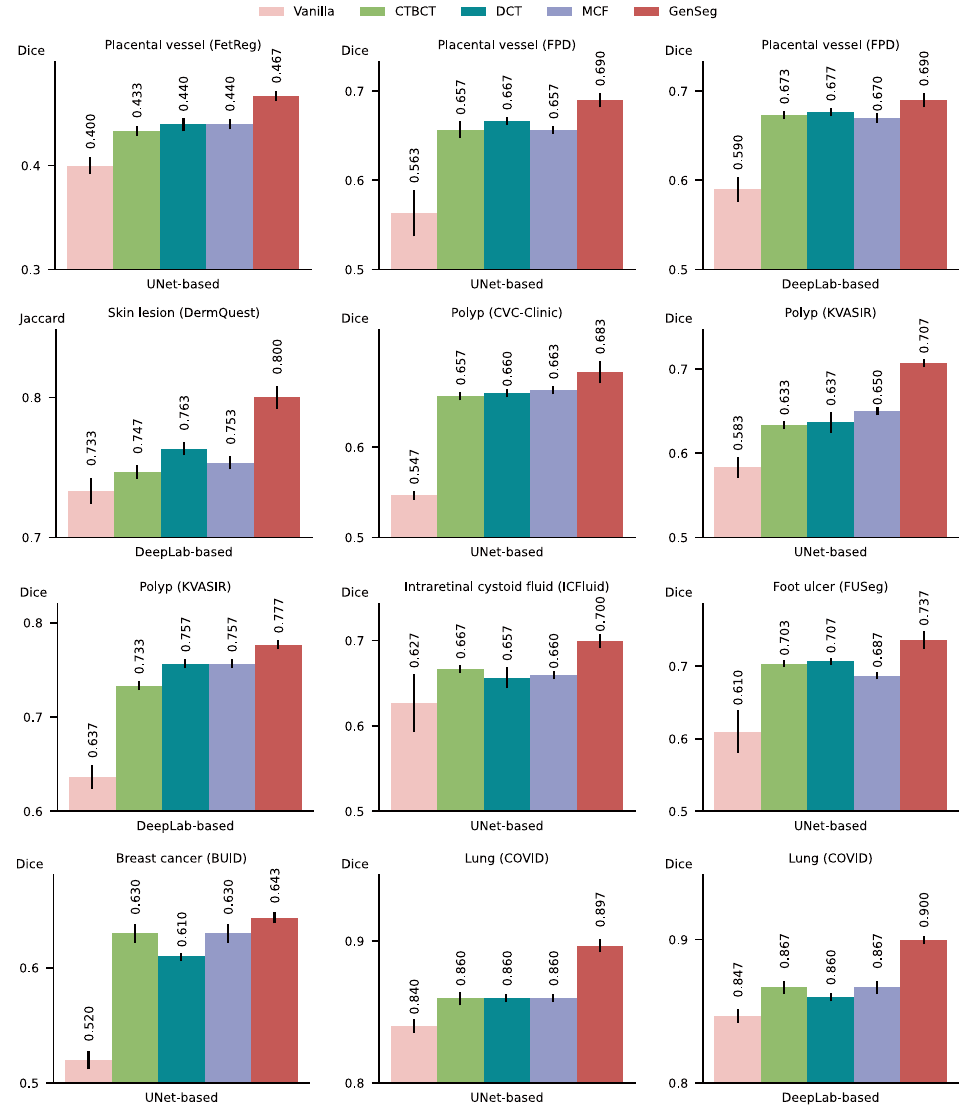}
    \captionof{supplementaryfigure}{\textbf{Further comparison of GegSeg with semi-supervised segmentation methods.} 
    GenSeg's in-domain generalization performance compared to baseline methods including CTBCT, DCT, and MCF, when used with UNet or DeepLab in segmenting placental vessels, skin lesions, polyps, intraretinal cystoid fluids, foot ulcers,  breast cancer, and lungs  utilizing 
    the FetReg, FPD, 
    DermQuest, CVC-Clinic, KVASIR, 
    ICFluid, FUSeg,  BUID, and COVID datasets.}
    \label{fig:semi_single}
\end{figure*}

\begin{figure*}
    \centering
    \includegraphics{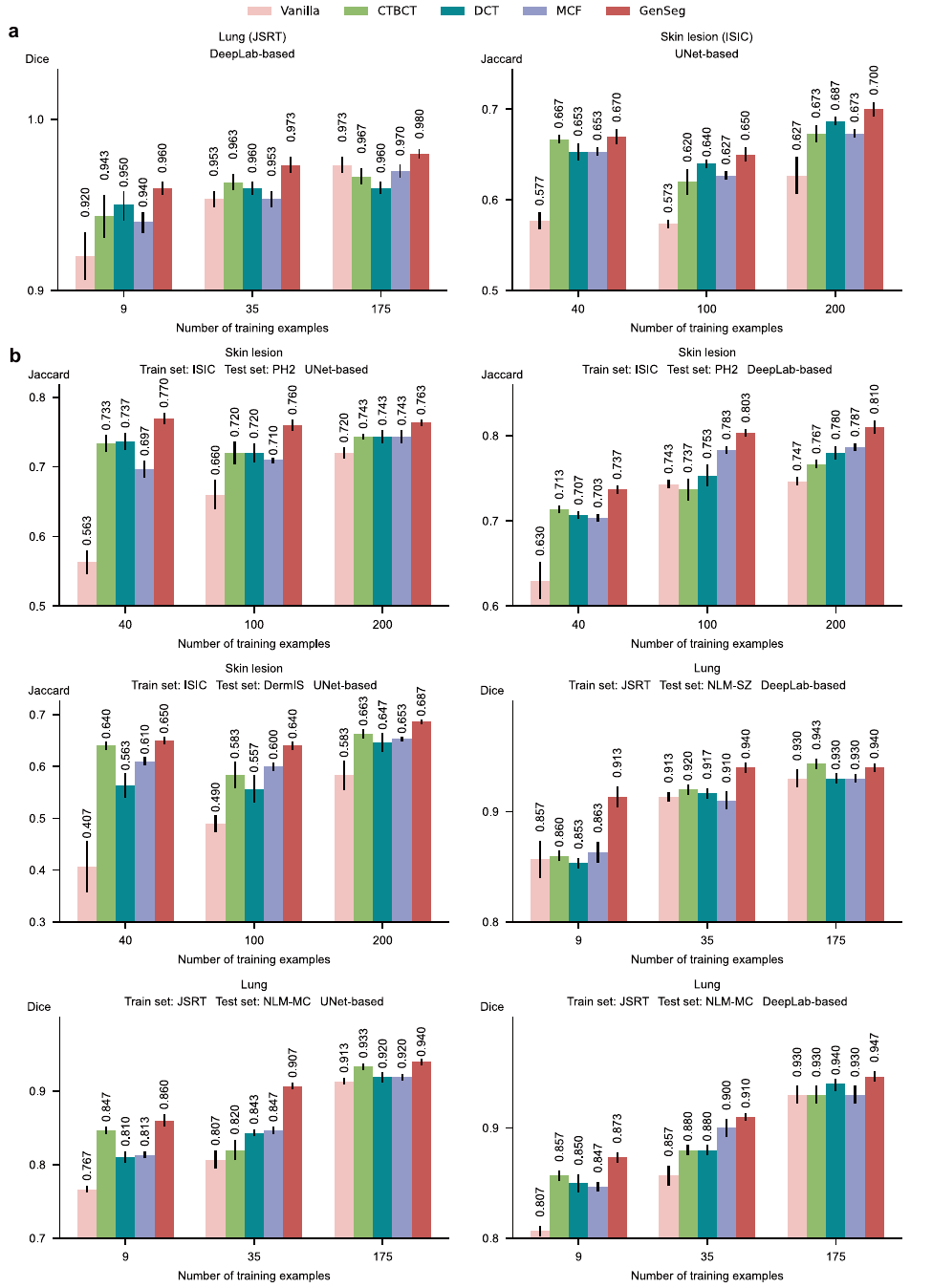}
    \captionof{supplementaryfigure}{\textbf{Further comparison of GegSeg with semi-supervised segmentation methods across varying numbers of training examples.} \textbf{a,} Comparison of in-domain generalization performance for segmenting lungs (using the JSRT dataset) and skin lesions (using ISIC). \textbf{b}, Comparison of out-of-domain generalization performance for segmenting skin lesions  (using ISIC for training, and PH2 and DermIS for testing) and lungs (using JSRT for training, and NLM-SZ and NLM-MC for testing).} 
    \label{fig:semi_varing}
\end{figure*}

We conducted a comparative analysis of GenSeg against leading semi-supervised segmentation methods~\cite{mendel2020semi,chen2021semi,peng2020deep,li2021semantic}, including cross-teaching between convolutional neural networks and Transformer (CTBCT)~\cite{luo2021ctbct}, deep co-training (DCT)~\cite{peng2020deep}, and a mutual correction framework (MCF)~\cite{wang2023mcf}, which  employ  external unlabeled images (1000  in each experiment) to enhance model training and thereby improve segmentation performance. GenSeg, which does not require any additional unlabeled images, significantly outperformed baseline methods under  in-domain settings (Fig.~\ref{fig:semi}a and Extended Data Fig. \ref{fig:semi_single}). For example, when using DeepLab as the backbone segmentation model for polyp segmentation, GenSeg achieved a Dice score of 0.76, markedly outperforming the top baseline method, MCF, which reached only 0.69. GenSeg also exhibited superior out-of-domain (OOD) generalization capabilities compared to baseline methods (Fig.~\ref{fig:semi}c and Extended Data Fig. \ref{fig:semi_varing}b). For instance, in skin lesion segmentation based on DeepLab with 40 training examples from the ISIC dataset, GenSeg achieved a Dice score of 0.67 on the DermIS dataset, significantly higher than the best-performing baseline, MCF, which scored 0.58. Additionally, GenSeg showed performance on par with baseline methods using fewer training examples in both in-domain (Fig.~\ref{fig:semi}b and Extended Data Fig. \ref{fig:semi_varing}a) and out-of-domain settings (Fig.~\ref{fig:semi}c and Extended Data Fig. \ref{fig:semi_varing}b).

\subsection*{GenSeg's end-to-end generation mechanism is superior to baselines' separate generation}

\begin{figure*}
    \centering
    \includegraphics{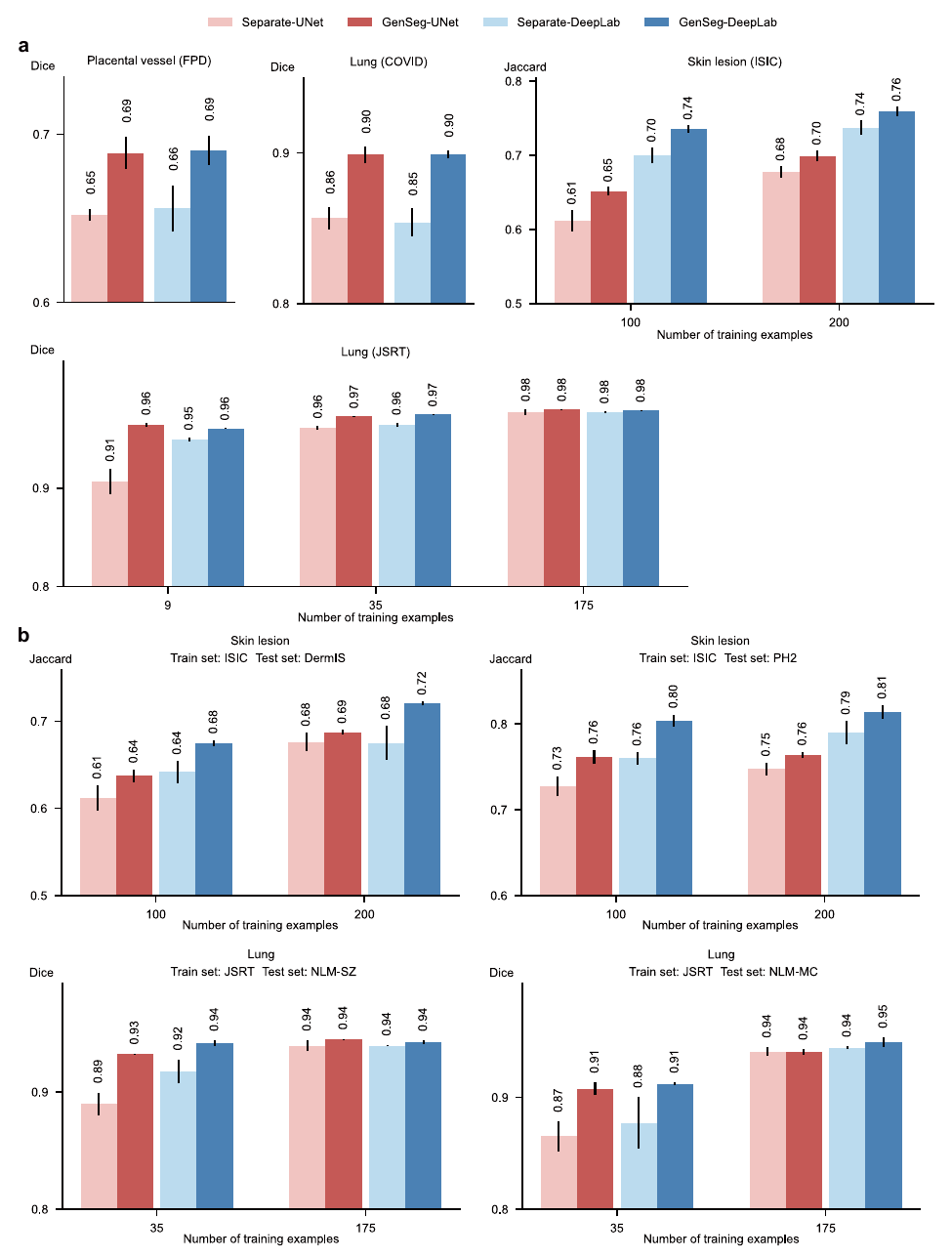}
    \captionof{supplementaryfigure}{\textbf{Further comparison of GenSeg's end-to-end data generation mechanism with baselines' separate generation mechanism.} \textbf{a,} GenSeg's end-to-end generation mechanism greatly improves models' in-domain generalization performance, when used UNet and DeepLab in segmenting placental vessels, lung regions, and skin lesions using FPD, COVID, ISIC, and JSRT datasets. \textbf{b,} GenSeg's end-to-end generation mechanism greatly improves models' out-of-domain generalization performance, when used UNet and DeepLab in segmenting skin lesions (using ISIC for training, and DermIS and PH2 for testing), and lung regions (using JSRT for training, and NLM-SZ and NLM-MC for testing).}
    \label{fig:sepa_appen}
\end{figure*}

We compared the effectiveness of GenSeg's end-to-end data generation mechanism against a baseline approach, Separate, which separates  data generation from segmentation model training. In Separate,  the mask-to-image generation model is initially trained and then fixed. Subsequently, it generates data, which is then utilized to train the segmentation model. The end-to-end GenSeg framework  consistently outperformed the Separate approach under both in-domain (Fig.~\ref{fig:resultend-to-end}a and Extended Data Fig. \ref{fig:sepa_appen}a) and out-of-domain settings (Fig.~\ref{fig:resultend-to-end}b and Extended Data Fig. \ref{fig:sepa_appen}b). For instance, in the segmentation of placental vessels,  GenSeg-DeepLab  attained an in-domain Dice score of 0.52, significantly surpassing  Separate-DeepLab, which scored 0.42.  In lung segmentation using JSRT as the training dataset,  GenSeg-UNet  achieved an out-of-domain Dice score of 0.93 on the NLM-SZ dataset, considerably better than the 0.84 scored by Separate-UNet.

\subsection*{GenSeg improves  the performance of diverse backbone segmentation models} 
\begin{figure*}[t]
    \centering
    \includegraphics{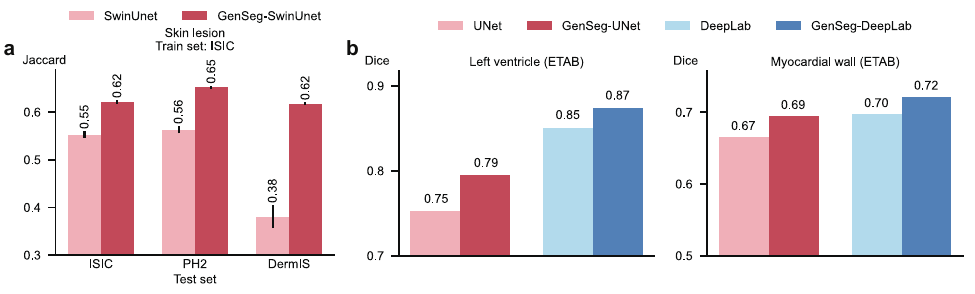}
    \captionof{supplementaryfigure}{
    \textbf{a}, Comparison between GenSeg-SwinUnet and SwinUnet models, both trained on 40 examples from the ISIC dataset and evaluated on the test sets of ISIC, PH2, and DermIS.
    \textbf{b}, The performance of GenSeg applied to UNet (GenSeg-UNet) and DeepLab (GenSeg-DeepLab) under in-domain settings (test and training data are from the same domain) in the tasks of segmenting left ventricles and myocardial wall using 8 training examples from the ETAB dataset, compared to vanilla UNet and DeepLab.}
    \label{fig:swin}
\end{figure*}

GenSeg is a versatile, model-agnostic framework that can seamlessly integrate with segmentation models with diverse architectures to improve their performance. 
After applying our framework on U-Net and DeepLab, we observed significant enhancements in their performance (Figs.~\ref{fig:main-results}-\ref{fig:resultend-to-end}), both for in-domain and out-of-domain  settings.  
Furthermore, we also integrated this framework with a Transformer-based segmentation model, SwinUnet~\cite{cao2022swin}. 
Using just 40 training examples from the ISIC dataset, GenSeg-SwinUnet achieved a Jaccard index of 0.62 on the ISIC test set. Furthermore, it demonstrated strong generalization with out-of-domain Jaccard index scores of 0.65 on the PH2 dataset and 0.62 on the DermIS dataset. These results represent a substantial improvement over the baseline SwinUnet model, which achieved Jaccard indices of 0.55 on ISIC, 0.56 on PH2, and 0.38 on DermIS  (Extended Data Fig.~\ref{fig:swin}a).

\section*{Discussion}
We present GenSeg, a generative deep learning framework designed for generating high-quality training data to enhance the training of medical image segmentation models. Demonstrating superior performance across  eight diverse segmentation tasks and 17 datasets, GenSeg excels particularly in scenarios with an extremely limited number of real, expert-annotated training examples (as few as 50). This ultra low-data regime often hinders the training of effective and broadly applicable segmentation models, especially those with hundreds of millions of parameters. GenSeg effectively overcomes this challenge by supplementing the training process with its generated high-fidelity data examples.
\newline
\newline
GenSeg stands out by requiring fewer expert-annotated real training examples compared to baseline methods, yet it achieves comparable performance. This substantial reduction in the need for manually labeled segmentation masks significantly cuts down both the burden and costs associated with medical image annotation. With just a small set of real examples, GenSeg effectively trains a data generation model which then produces additional synthetic data, effectively mimicking the benefits of using a large dataset of real examples.
\newline
\newline
GenSeg significantly improves segmentation models' out-of-domain (OOD) generalization capability. GenSeg is capable of generating diverse medical images accompanied by precise segmentation masks. When trained on this diverse augmented dataset, segmentation models can learn more robust and OOD generalizable feature representations.  
\newline
\newline
GenSeg stands out from current data augmentation and generation techniques by offering superior segmentation performance, primarily due to its end-to-end data generation mechanism. Unlike previous methods that separate data augmentation/generation and segmentation model training, our approach integrates them end-to-end within a unified, multi-level optimization framework. Within this framework,  the validation performance of the segmentation model acts as a direct indicator of the generated data's usefulness. 
By leveraging this performance to inform the training process of the generation model, we ensure that the data produced is specifically optimized to improve the segmentation model.  In previous methods,    segmentation performance does not impact the process of data augmentation and generation. As a result,   the augmented/generated data might not be effectively tailored for training the segmentation model.
Furthermore, 
our framework learns a generative model that excels in generating data with greater diversity compared to existing augmentation methods. 
\newline
\newline
GenSeg excels in surpassing semi-supervised segmentation methods without the need for external unlabeled images.  
In the context of medical imaging, collecting even unlabeled images presents a significant challenge due to stringent privacy concerns and regulatory constraints (e.g., IRB approval), thereby reducing the feasibility of semi-supervised methods. Despite the use of unlabeled real images, semi-supervised approaches underperform compared to GenSeg. This is primarily because these methods struggle to generate accurate masks for unlabeled images, meaning they are less effective at creating labeled training data. On the other hand, GenSeg is capable of producing high-quality images from masks, ensuring a close correspondence between the images' content and the masks, thereby efficiently generating labeled training examples.
\newline
\newline
Our framework is designed to be universally applicable and independent of specific models. This design choice enables it to augment the capabilities of a broad spectrum of semantic segmentation models. To apply our framework to a specific segmentation model, the only requirement is to integrate the segmentation model into the second and third stages of our framework. This straightforward process enables researchers and practitioners to easily utilize our approach to improve the performance of diverse semantic segmentation models. 
\newline
\newline
In summary, GenSeg is a robust data generation tool that seamlessly integrates with current semantic segmentation models. It significantly enhances both in-domain and out-of-domain generalization performance in ultra low-data regimes, markedly boosting sample efficiency. Furthermore, it surpasses state-of-the-art methods in data augmentation and semi-supervised learning.

\section*{Methods}

\subsection*{Overview of GenSeg}
GenSeg consists of a data generation model and a medical image segmentation model. The data generation model is based on     conditional generative adversarial networks (GANs)~\citeMethod{mirza2014conditional,Isola_2017_CVPR}. It  comprises two main components: a mask-to-image generator and a discriminator. Uniquely, our generator has a learnable neural architecture~\citeMethod{liudarts}, as opposed to the fixed architecture commonly seen in previous GAN models. This generator, with weight parameters $G$ and a learnable architecture $A$, takes a segmentation mask as input and generates a corresponding medical image. The discriminator, with learnable weight parameters $H$ and a fixed architecture, differentiates between synthetic and real medical images. The segmentation model has learnable weight parameters $S$ and a fixed architecture. 
\newline
\newline
Data generation is executed in a reverse manner. Starting with an expert-annotated segmentation mask $M$, we first apply basic image augmentations, such as rotation, flipping, etc., to produce an augmented mask $\widehat{M}$.  This mask is then fed into the mask-to-image  generator, resulting in a medical image $\hat{I}(\widehat{M},G,A)$, which corresponds to $\widehat{M}$, i.e.,  pixels in $\hat{I}(\widehat{M},G,A)$ can be semantically labeled using $\widehat{M}$. 
Each image-mask pair $(\hat{I}(\widehat{M},G,A),\widehat{M}) $ forms an augmented example for training the segmentation model. Like other deep learning-based segmentation methods, GenSeg has access to a training set comprised of real image-mask pairs $D^{tr}_{seg}=\{I^{(tr)}_n, M^{(tr)}_n\}_{n=1}^{N_{tr}}$ and a validation set $D^{val}_{seg}=\{I^{(val)}_n, M^{(val)}_n\}_{n=1}^{N_{val}}$.

\subsection*{A multi-level optimization framework for GenSeg}
GenSeg employs a multi-level optimization strategy across three distinct stages. The initial stage focuses on training the data generation model, where we fix the generator's architecture $A$ and train the weight parameters of both the generator ($G$) and the discriminator ($H$). To facilitate this training, we modify the  segmentation training dataset $D^{tr}_{seg}$ by swapping the roles of inputs and outputs, resulting in a new dataset $D_{gan}=\{M^{(tr)}_n,I^{(tr)}_n\}_{n=1}^{N_{tr}}$. In this setup, $M^{(tr)}_n$ serves as the input, while $I^{(tr)}_n$ acts as the output for our mask-to-image GAN model. 
\newline
\newline
Let $L_{gan}$ represent the GAN training objective, a cross-entropy function that evaluates the discriminator's ability to distinguish between real and generated images. The discriminator's goal is to maximize $L_{gan}$, effectively separating real images from generated ones. Conversely, the generator strives to minimize $L_{gan}$, generating images that are so realistic they become indistinguishable from real ones. This process is encapsulated in the following  minimax optimization problem:
\begin{equation}
    G^*(A), H^* = \underset{G}{\textrm{argmin}} \, \underset{H}{\textrm{argmax}} \,\; L_{gan}(G, A, H, D_{gan}),
\end{equation}
where $G^*(A)$ indicates that the optimally trained generator $G^*$ is dependent on the architecture $A$. This dependency arises because $G^*$ is the outcome of optimizing the training objective function, which in turn is influenced by $A$. $A$ is tentatively fixed at this stage and will be updated later. Otherwise, if we learn  $A$ by minimizing the training loss $L_{gan}$, it may lead to a trivial solution characterized by an overly large and complex $A$. Such a solution would likely fit the training data perfectly but perform inadequately on unseen test data due to overfitting. 
\newline
\newline
In the second stage, we leverage the trained generator to generate synthetic training examples  using the aforementioned process where expert-annotated masks are  from $D^{tr}_{seg}$. Let $\widehat{D}(G^*(A),D^{tr}_{seg})$ represent the  generated data. We then use $\widehat{D}(G^*(A),D^{tr}_{seg})$ and  real training data $D^{tr}_{seg}$  to train the segmentation model $S$ by minimizing a segmentation loss $L_{seg}$ (pixel-wise
cross-entropy loss). This training is formulated as the following optimization problem:
{\small
\begin{equation} \label{eq: stage2}
   S^*(A) = \underset{S}{\textrm{argmin}} \,  L_{seg}(S, \widehat{D}(G^*(A),D^{tr}_{seg})) + \gamma L_{seg}(S, D^{tr}_{seg}), 
\end{equation} 
}
where $\gamma$ is a trade-off parameter.
\newline
\newline
In the third stage, we assess the performance of the trained segmentation model on the validation dataset $D^{val}_{seg}$. The validation loss, $L_{seg}(S^*(A), D^{val}_{seg})$, serves as an indicator of the quality of the generated data. If the generated data is of inferior quality, it will likely result in $S^*(A)$ - trained on this data - performing poorly on the validation set, reflected in a high validation loss. Thus, enhancing the quality of generated data can be achieved by minimizing $L_{seg}(S^*(A), D^{val}_{seg})$ w.r.t the generator's architecture $A$. This objective is encapsulated in the following optimization problem:
\begin{equation}
    \underset{A}{\textrm{min}} \, L_{seg}(S^*(A), D^{val}_{seg}).
\end{equation}
\newline
We can integrate these stages into a multi-level optimization problem as follows: 
\begin{align}\label{eq:mlo}
\textrm{min}_A &\quad L_{seg}(S^*(A), D^{val}_{seg}) \nonumber \\
s . t &\quad S^*(A) =\underset{S}{\textrm{argmin}} \,  L_{seg}(S, \widehat{D}(G^*(A),D^{tr}_{seg})) + \nonumber \\
   &\qquad\qquad\qquad\qquad\qquad\qquad\quad \gamma L_{seg}(S, D^{tr}_{seg}) \\
&\quad G^*(A), H^*=\underset{G}{\textrm{argmin}} \, \underset{H} {\textrm{argmax}} \,\; L_{gan}(G, A, H, D_{gan})  \nonumber
\end{align}

In this formulation, the levels are interdependent. The output $G^*(A)$ from the first level defines the objective for the second level, the output $S^*(A)$ from the second level defines the objective for the third level, and the optimization variable $A$ in the third level defines the objective function in the first level.

\subsection*{Architecture search space}

To enhance the generation of medical images by accurately capturing their distinctive characteristics, we make the generator's architecture searchable.  Inspired by DARTS~\citeMethod{liu2018darts}, we employ a differentiable search method that is not only computationally efficient but also allows for a flexible exploration of architectural designs. Our search space is structured as a series of computational cells, each forming a directed acyclic graph that includes an input node, an output node, and intermediate nodes comprising $K$ different operators, such as convolution and transposed convolution. These operators are each tied to a learnable selection weight, $\alpha$, ranging from 0 to 1, where a higher $\alpha$ value indicates a stronger preference for incorporating that operator into the final architecture. The process of architecture search is essentially the optimization of these selection weights. Let Conv-$xyz$ and UpConv-$xyz$ denote a convolution operator and a transposed convolution operator respectively, where $x$ represents the kernel size, $y$ the stride, and $z$ the padding. The pool of candidate operators includes Conv/UpConv-421, Conv/UpConv-622, and Conv/UpConv-823, i.e., the number of operators $K$ is 3. 
For any given cell $i$ with input $x_i$, the output $y_i$ is determined by the formula $y_i = \sum_{k=1}^K \alpha_{i, k} \boldsymbol{o}_{i, k}(x_i)$, where $\boldsymbol{o}_{i, k}$ represents the $k$-th operator in the cell, and $\alpha_{i, k}$ is its corresponding selection weight. Consequently, the architecture of the generator can be succinctly described by the set of all selection weights, denoted as $A=\{\alpha_{i, k}\}$. Architecture search amounts to learning $A$. 

\subsection*{Optimization algorithm}
We develop a gradient-based method to solve the multi-level optimization problem in Eq.(\ref{eq:mlo}). First, we approximate $G^*(A)$ using one-step gradient descent update of $G$ w.r.t $L_{gan}(G, A, H, D_{gan})$: 
\begin{equation}
\label{eq:update_g}
    G^*(A)\approx G'= G-\eta_{g} \nabla_G  L_{gan}(G, A, H, D_{gan}),
\end{equation}
where $\eta_{g}$ is a learning rate. 
Similarly, we approximate $H^*$ using one-step gradient ascent update of $H$ w.r.t $L_{gan}(G, A, H, D_{gan})$:
\begin{equation}
    H^*\approx H'= H+\eta_{h} \nabla_H L_{gan}(G, A, H, D_{gan}).
\end{equation}
Then we plug $G^*(A)\approx G'$ into the objective function in the second level, yielding an approximated objective. We approximate $S^*(A)$ using one-step gradient ascent update of $S$ w.r.t the approximated objective:
\begin{align}
    S^*(A)\approx S'=S-\eta_{s} \nabla_S(L_{seg}(&S, \widehat{D}(G',D^{tr}_{seg})) + \nonumber \\
    &\gamma L_{seg}(S, D^{tr}_{seg})). 
\end{align}
Finally, we plug $S^*(A)\approx S'$ into the validation loss in the third level, yielding an approximated validation loss. We update $A$ using gradient descent w.r.t the approximated loss:
\begin{equation}
\label{eq:update_a}
    A\gets A-\eta_{a}\nabla_A L_{seg}(S', D^{val}_{seg}). 
\end{equation}
After $A$ is updated, we plug it into Eq.(\ref{eq:update_g}) to update $G$ again. The update steps in Eq.(\ref{eq:update_g}-\ref{eq:update_a}) iterate until convergence. 
\newline
The gradient $\nabla_A L_{seg}(S', D^{val}_{seg})$ can be calculated as follows:
\begin{equation}
    \nabla_A L_{seg}(S', D^{val}_{seg})= 
    \frac{\partial G'}{\partial A}
    \frac{\partial S'}{\partial G'}
    \frac{\partial L_{seg}(S', D^{val}_{seg})}{\partial S'}, 
\end{equation}
where 
\begin{equation}
    \frac{\partial G'}{\partial A}= -\eta_{g} \nabla^2_{A,G}  L_{gan}(G, A, H, D_{gan}),
\end{equation}
\begin{align}
    \frac{\partial S'}{\partial G'}=
    -\eta_{s} \nabla^2_{G',S}(L_{seg}(S, \widehat{D}&(G',D^{tr}_{seg})) + \nonumber \\
    \gamma &L_{seg}(S, D^{tr}_{seg})). 
\end{align}

\begin{table}[]
    \centering
    \footnotesize
    \begin{tabular}{c|c|ccc}
        \toprule
        Task & Dataset & Train & Validate & Test \\ \hline\hline
        \multirow{4}{*}{\makecell{Skin lesion \\ segmentation}} & ISIC & 160 & 40 & 594 \\
        & PH2 & - & - & 200 \\
        & DermIS & - & - & 98 \\
        & DermQuest & 32 & 8 & 61 \\ \hline
        \multirow{4}{*}{\makecell{Lung \\ segmentation}} & JSRT & 140 & 35 & 72 \\
        & NLM-MC & - & - & 138 \\
        & NLM-SZ & - & - & 566 \\
        & COVID & 8 & 2 & 583 \\ \hline
        \makecell{Breast cancer \\ segmentation} & BUID & 80 & 20 & 230 \\ \hline
        \multirow{2}{*}{\makecell{Placental vessel \\ segmentation}} & FPD & 80 & 20 & 182 \\
        & FetReg & 80 & 20 & 658 \\ \hline
            \multirow{2}{*}{\makecell{Polyp \\ segmentation}} & KVASIR & 480 & 120 & 200 \\
        & CVC-Clinic & 80 & 20 & 212 \\ \hline
        \makecell{Foot ulcer \\ segmentation} & FUSeg & 480 & 120 & 200 \\ \hline
        \makecell{Intraretinal cystoid \\ segmentation} & ICFluid & 40 & 10 & 460 \\ \hline
        \makecell{Left ventricle \\ segmentation} & \makecell{ETAB\\(Left ventricle)} & 8 & 2 & 50 \\ \hline
        \makecell{Myocardial wall \\ segmentation} & \makecell{ETAB\\(Myocardial wall)} & 8 & 2 & 50 \\
        \bottomrule
    \end{tabular}
    \caption{Dataset statistics.}
    \label{tab:other-data1}
\end{table}

\subsection*{Datasets}
In this study, we focused on the segmentation of skin lesions from dermoscopy images, lungs from chest X-ray images, breast cancer from ultrasound images, placental vessels from fetoscopic images, polyps from colonoscopy images, foot ulcers from standard camera images,  intraretinal cystoid fluid from optical coherence tomography (OCT) images, and left ventricle and myocardial wall from echocardiography images, utilizing 16 datasets.  Each dataset was randomly partitioned into training, validation, and test sets, with the corresponding statistics presented in Table~\ref{tab:other-data1}.
\newline
\newline
For skin lesion segmentation from dermoscopy images, we utilized the ISIC2018~\citeMethod{codella2019skin}, PH2~\citeMethod{mendoncca2013ph},  DermIS~\citeMethod{glaister2013automatic}, and DermQuest~\citeMethod{chung2015statistical} datasets. The ISIC2018 dataset, provided by the International Skin Imaging Collaboration (ISIC) 2018 Challenge, comprises 2,594 dermoscopy images, each meticulously annotated with pixel-level skin lesion labels. The PH2 dataset, acquired at the Dermatology Service of Hospital Pedro Hispano in Matosinhos, Portugal, contains 200 dermoscopic images of melanocytic lesions. These images are in 8-bit RGB color format with a resolution of 768x560 pixels. DermIS offers a comprehensive collection of dermatological images covering a range of skin conditions, including dermatitis, psoriasis, eczema, and skin cancer. DermQuest includes 137 images representing two types of skin lesions: melanoma and nevus. 
\newline
\newline 
For lung segmentation from chest X-rays, we utilized the JSRT~\citeMethod{shiraishi2000development}, NLM-MC~\citeMethod{jaeger2014two}, NLM-SZ~\citeMethod{jaeger2014two}, and COVID-QU-Ex~\citeMethod{covid_kaggle} datasets. 
The JSRT dataset consists of 247 chest X-ray images from Japanese patients, each accompanied by manually annotated ground truth masks that delineate the lung regions. 
The NLM-MC dataset was collected from the Department of Health and Human Services in Montgomery County, Maryland, USA. It includes 138 frontal chest X-rays, with manual lung segmentations provided. Of these, 80 images represent normal cases, while 58 exhibit manifestations of tuberculosis (TB). The images are available in two resolutions: 4,020x4,892 pixels and 4,892x4,020 pixels. 
The NLM-SZ dataset, sourced from Shenzhen No.3 People's Hospital, Guangdong, China, contains 566 frontal chest X-rays in PNG format. Image sizes vary but are approximately 3,000x3,000 pixels. 
The COVID-QU-Ex dataset, compiled by researchers at Qatar University, comprises a large collection of chest X-ray images, including 11,956 COVID-19 cases, 11,263 non-COVID infections, and 10,701 normal instances. Ground-truth lung segmentation masks are provided for all images in this dataset.   
\newline
\newline
For placental vessel segmentation from fetoscopic images, we utilized the FPD~\citeMethod{bano2020vessel} and FetReg~\citeMethod{bano2021fetreg} datasets. The FPD dataset comprises 482 frames extracted from six distinct in vivo fetoscopic procedure videos. To reduce redundancy and ensure a diverse set of annotated samples, the videos were down-sampled from 25 to 1 fps, and each frame was resized to a resolution of 448x448 pixels. Each frame is provided with a corresponding segmentation mask that precisely outlines the blood vessels. 
The FetReg dataset, developed for the FetReg2021 challenge, is the first large-scale, multi-center dataset focused on fetoscopy laser photocoagulation procedures. It contains 2,718 pixel-wise annotated images, categorizing background, vessel, fetus, and tool classes, sourced from 24 different in vivo TTTS fetoscopic surgeries.   
\newline
\newline
For polyp segmentation from colonoscopic images, we utilized the KVASIR~\citeMethod{jha2020kvasir} and CVC-ClinicDB~\citeMethod{bernal2015wm} datasets. Polyps are recognized as precursors to colorectal cancer and are detected in nearly half of individuals aged 50 and older who undergo screening colonoscopy, with their prevalence increasing with age. Early detection of polyps significantly improves survival rates from colorectal cancer. 
The KVASIR dataset was collected using endoscopic equipment at Vestre Viken Health Trust (VV) in Norway, which consists of four hospitals and provides healthcare services to a population of 470,000. The dataset includes images with varying resolutions, ranging from 720x576 to 1920x1072 pixels. It contains 1,000 polyp images, each accompanied by a corresponding segmentation mask, with annotations verified by experienced endoscopists. 
CVC-ClinicDB comprises frames extracted from colonoscopy videos and consists of 612 images with a resolution of 384x288 pixels, derived from 31 colonoscopy sequences. videos. 
\newline
\newline
For breast cancer segmentation, we utilized the BUID dataset~\citeMethod{al2020dataset}, which consists of 630 breast ultrasound images collected from 600 female patients aged between 25 and 75 years. The images have an average resolution of 500x500 pixels. 
For foot ulcer segmentation, we utilized data from the FUSeg challenge~\citeMethod{wang2020fully}, which includes over 1,000 images collected over a span of two years from hundreds of patients. The raw images were captured using  Canon SX 620 HS digital cameras and iPad Pro under uncontrolled lighting conditions, with diverse backgrounds. 
For the segmentation of intraretinal cystoids from  Optical Coherence Tomography (OCT) images, we utilized the Intraretinal Cystoid Fluid (ICFluid) dataset~\citeMethod{zeeshan2022}. This dataset comprises 1,460 OCT images along with their corresponding masks for the Cystoid Macular Edema (CME) ocular condition. 
For the segmentation of  left ventricles and myocardial wall, we employed data examples from the ETAB benchmark~\citeMethod{m2022etab}. It  is constructed from five publicly available echocardiogram datasets, encompassing diverse cohorts and providing echocardiographies with a variety of views and annotations.

\subsection*{Metrics}

For all segmentation tasks except skin lesion segmentation, we used the Dice score as the evaluation metric, adhering to established conventions in the field~\citeMethod{bertels2019optimizing}. The Dice score is calculated as $\frac{2|A \cap B|}{|A| + |B|}$, where $A$ represents the algorithm's prediction and $B$ denotes the ground truth. For skin lesion segmentation, we followed the guidelines of the ISIC challenge~\citeMethod{rotemberg2021patient} and employed the Jaccard index, also known as intersection-over-union (IoU), as the performance metric. The Jaccard index is computed as $\frac{|A \cap B|}{|A \cup B|}$ for each patient case. These metrics provide a robust assessment of the overlap between the predicted segmentation mask and the ground truth.

\subsection*{Hyperparameters}

In our method, mask augmentation was performed using a series of operations, including rotation, flipping, and translation, applied in a random sequence. The mask-to-image generation model was based on the Pix2Pix framework~\citeMethod{Isola_2017_CVPR}, with an architecture that was made searchable, as depicted in Fig.~\ref{fig:overview}b.  The tradeoff parameter $\gamma$ was set to 1.  
We configured the training process to perform 5,000 iterations. The RMSprop optimizer~\citeMethod{shi2021rmsprop} was utilized for training the segmentation model. It was set with an initial learning rate of $1e-5$, a momentum of 0.9, and a weight decay of $1e-3$. Additionally, the ReduceLROnPlateau scheduler was employed to dynamically adjust the learning rate according to the model's performance throughout the training period.  Specifically, the scheduler was configured with a patience of 2 and set to `max' mode, meaning it monitored the model's validation performance and adjusted the learning rate to maximize validation accuracy.  
For training the mask-to-image generation model, the Adam optimizer~\citeMethod{kingma2014adam}  was chosen, configured with an initial learning rate of $1e-5$, beta values of (0.5, 0.999), and a weight decay of $1e-3$. Adam was also applied for optimizing the  architecture variables, with a learning rate of $1e-4$, beta values of (0.5, 0.999), and weight decay of $1e-5$. 
At the end of each epoch, we assessed the performance of the trained segmentation model on a validation set. The model checkpoint with the best validation performance was selected as the final model. 
The experiments were conducted on A100 GPUs, with each method being run three times using randomly initialized model weights. We report the average results along with the standard deviation across these three runs.

\subsection*{The impact of  the tradeoff parameter $\lambda$ on segmentation performance}

We investigated the effect of the hyperparameter $\lambda$ in Eq.(\ref{eq: stage2}) on the performance of our method. This parameter controls the balance between the contributions of real and generated  data during the training of the segmentation model. Optimal performance was observed with a moderate $\lambda$ value (e.g., 1), which effectively balanced the use of real and generated  data (Extended Data Fig.~\ref{fig:ablation_lambda_mask}a).

\subsection*{The impact of mask augmentation operations on segmentation performance}

In GenSeg, the initial step involves applying augmentation operations to generate synthetic segmentation masks from real masks. We explored the impact of augmentation operations on segmentation performance. GenSeg, which utilizes all three operations - rotation, translation, and flipping - is compared against three specific ablation settings where only one operation (Rotate, Translate, or Flip) is used to augment the masks. 
 GenSeg demonstrated significantly superior performance compared to any of the individual ablation settings (Extended Data Fig.~\ref{fig:ablation_lambda_mask}b). Notably, GenSeg  exhibited superior generalization on out-of-domain data, highlighting the advantages of integrating multiple augmentation operations compared to using a  single operation. By combining various augmentation operations, GenSeg can  generate a broader diversity of augmented masks, which in turn produces a more diverse set of augmented images. Training segmentation models on this diverse  dataset allows for  learning  more robust representations, thereby significantly enhancing  generalization capabilities on out-of-domain test data.

\subsection*{The impact of  mask-to-image GANs on segmentation performance} 

We investigated the impact of the mask-to-image conditional Generative Adversarial Network (GAN) in GegSeg on segmentation performance by comparing the default Pix2Pix model with two other conditional GAN models: SPADE~\citeMethod{park2019SPADE} and ASAPNet~\citeMethod{shaham2021spatially}. In this comparison, we made the  architectures of these models' generators  searchable. Pix2Pix and SPADE demonstrated comparable performance, both significantly outperforming ASAPNet (Extended Data Fig.~\ref{fig:ablation_lambda_mask}c). This performance gap can be attributed to the superior image generation capabilities of Pix2Pix and SPADE.

\subsection*{Computation costs}

\begin{figure*}[t]
    \centering
    \includegraphics[]{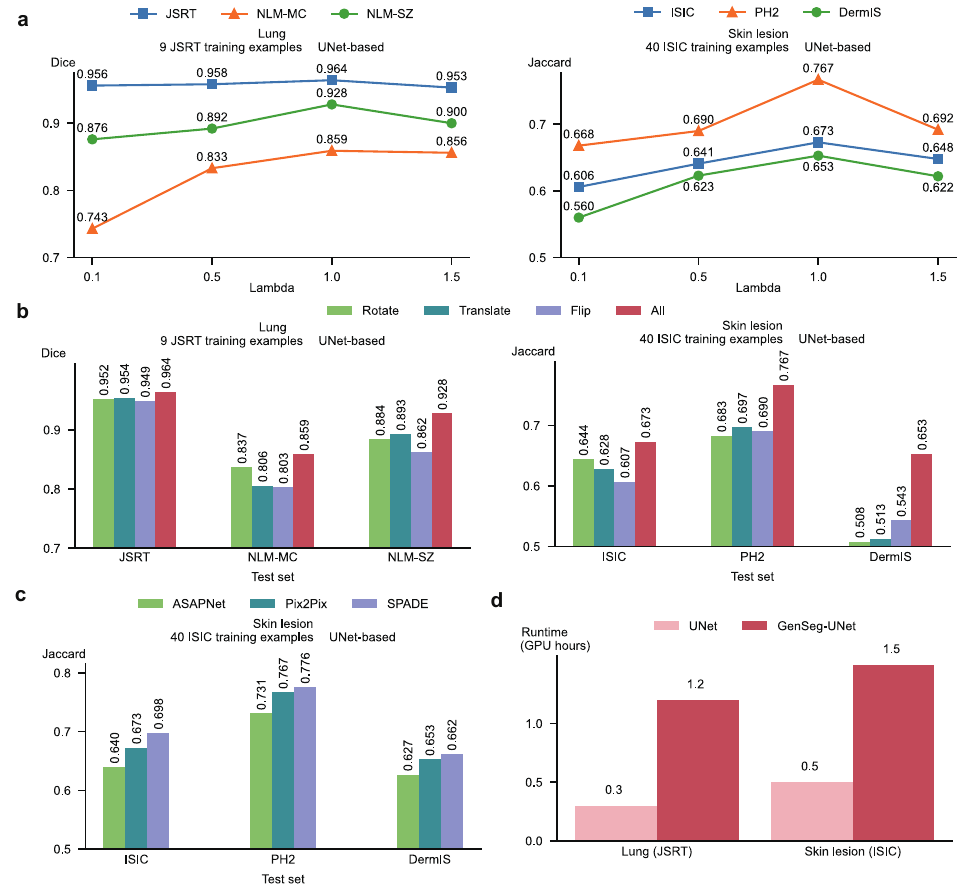}
    \captionof{supplementaryfigure}{\textbf{a}, (Left) Impact of the tradeoff parameter \(\lambda\) on the performance of GenSeg-UNet was evaluated on the test datasets of JSRT, NLM-MC, and NLM-SZ, in lung segmentation. GenSeg-UNet was trained using 9 examples from the JSRT training dataset. (Right) Impact of the tradeoff parameter \(\lambda\) on the performance of GenSeg-UNet was evaluated on the test datasets of ISIC, PH2, and DermIS, in skin lesion segmentation. GenSeg-UNet was trained using 40 examples from the ISIC  training dataset. \textbf{b}, (Left) Impact of augmentation operations  on the performance of GenSeg-UNet was evaluated on the test datasets of JSRT, NLM-MC, and NLM-SZ, in lung segmentation. GenSeg-UNet was trained using 9 examples from the JSRT training dataset. \textit{All} refers to the full GenSeg method that incorporates all three operations. (Right) Impact of augmentation operations  on the performance of GenSeg-UNet was evaluated on the test datasets of ISIC, PH2, and DermIS, in skin lesion segmentation. GenSeg-UNet was trained using 40 examples from the ISIC  training dataset.  
    \textbf{c}, Impact of mask-to-image GAN models  on the performance of GenSeg-UNet was evaluated on the test datasets of ISIC, PH2, and DermIS, in skin lesion segmentation. GenSeg-UNet was trained using 40 examples from the ISIC  training dataset.
    \textbf{d}, The runtime (in hours on an A100 GPU) of GenSeg-UNet was measured for lung segmentation using JSRT as the training data and for skin lesion segmentation using ISIC as the training data.}
\label{fig:ablation_lambda_mask}
\end{figure*}

Given that GenSeg is designed for scenarios with limited training data, the overall training time is minimal, often requiring less than 2 GPU hours   (Extended Data Fig.~\ref{fig:ablation_lambda_mask}d). To enhance the efficiency of GenSeg's training, we plan to incorporate strategies from~\citeMethod{sinha2020small,sinha2020top} for accelerated GAN training and implement the algorithm proposed in~\citeMethod{sato2021gradient} to expedite the convergence of multi-level optimization. Importantly, our method does not increase the inference cost of the segmentation model. This is because our approach maintains the original architecture of the segmentation model, ensuring that the Multiply-Accumulate (MAC) operations remain unchanged.

\section*{Data availability}
 
Datasets used in this study are available at  \href{https://challenge.isic-archive.com/data/}{\textit{ISIC}}, \href{https://www.fc.up.pt/addi/ph2%20database.html}{\textit{PH2}}, \href{https://uwaterloo.ca/vision-image-processing-lab/research-demos/skin-cancer-detection}{\textit{DermIS and DermQuest}}, \href{http://db.jsrt.or.jp/eng.php}{\textit{JSRT}}, \href{http://archive.nlm.nih.gov/repos/chestImages.php}{\textit{NLM-MC and NLM-SZ}}, \href{https://www.kaggle.com/datasets/anasmohammedtahir/covidqu}{\textit{COVID-QU-Ex Dataset}}, \href{https://www.kaggle.com/datasets/aryashah2k/breast-ultrasound-images-dataset?select=Dataset_BUSI_with_GT}{\textit{BUID}}, \href{https://www.ucl.ac.uk/interventional-surgical-sciences/fetoscopy-placenta-data}{\textit{FPD}}, \href{https://www.ucl.ac.uk/interventional-surgical-sciences/weiss-open-research/weiss-open-data-server/fetreg-largescale-multi-centre-fetoscopy-placenta-dataset}{\textit{FetReg}}, \href{https://datasets.simula.no/kvasir/}{\textit{KVASIR}}, \href{https://www.kaggle.com/datasets/balraj98/cvcclinicdb}{\textit{CVC-Clinic}}, \href{https://github.com/uwm-bigdata/wound-segmentation/tree/master}{FUSed},    
\href{https://www.kaggle.com/datasets/zeeshanahmed13/intraretinal-cystoid-fluid}{ICFluid}, and \href{https://github.com/AlaaLab/ETAB/tree/main}{ETAB}.

\section*{Code availability}
Our GenSeg code is available in the GitHub repository \url{https://github.com/importZL/semantic_segmentation}.

\section*{References}
\bibliographyMethod{Article}

\end{document}